\documentclass[journal,twocolumn]{IEEEtran}
\usepackage{color}
\usepackage{graphicx}
\usepackage{epstopdf}
\usepackage{amsmath}
\usepackage{amssymb}
\usepackage{hyperref}
\usepackage[english]{babel}
\usepackage{cite}
\usepackage{rotfloat}
\usepackage{mathtools}
\usepackage{amsmath}
\usepackage{makecell}
\usepackage{algorithm,algorithmic}
\usepackage{multirow}
\usepackage{subfigure}
\usepackage{booktabs}
\usepackage{colortbl}
\usepackage{multirow}% http://ctan.org/pkg/multirow
\usepackage{hhline}% http://ctan.org/pkg/hhline
\usepackage{stfloats}% <-- added
\usepackage{multicol}
\usepackage[justification=centering,font=small,skip=0pt]{caption}

\newsavebox{\foobox}

\newcommand{\setA}{\mathbb{A}}

\definecolor{kugray5}{RGB}{224,224,224}

\usepackage[normalem]{ulem}
\newcommand\rsout{\bgroup\markoverwith
	{\textcolor{red}{\rule[0.5ex]{2pt}{0.8pt}}}\ULon}



\makeatletter
\newcommand{\ALOOP}[1]{\ALC@it\algorithmicloop\ #1%
	\begin{ALC@loop}}
	\newcommand{\ENDALOOP}{\end{ALC@loop}\ALC@it\algorithmicendloop}

\makeatother

\usepackage{etoolbox}
\let\mybibitem\bibitem
\renewcommand{\bibitem}[1]{%
	\ifstrequal{#1}{nature}
	{\color{blue}\mybibitem{#1}}
	{\color{black}\mybibitem{#1}}%
}

\newtheorem{remark}{Remark}

\DeclareCaptionLabelSeparator{periodspace}{.\quad}

\addto\captionsenglish{}
\newcommand\nbthis{\addtocounter{equation}{1}\tag{\theequation}}

\newcommand{\norm}[1]{\left\lVert#1\right\rVert} % ||.||
\newcommand{\abs}[1]{\left|#1\right|} % ||

\newcommand{\tr}[1]{\mathrm{trace}\left(#1\right)} % ||
\newcommand{\diag}[1]{\mathrm{diag}\left\{#1\right\}} % ||

\newcommand{\re}[1]{\mathfrak{R}{\left(#1\right)}}

\newcommand{\mean}[1]{\mathbb{E} \left\{#1\right\}}

\newcommand{\phybrid}{\left(\mathrm{P}_{\mathrm{hybrid}} \right)}

\newcommand{\pdyn}{\left(\mathrm{P}_{\mathrm{dyn}} \right)}
\newcommand{\pdyna}{\left(\mathrm{P1}_{\mathrm{dyn}} \right)}

\newcommand{\pupdate}{\left(\mathrm{P_{update}}\right)}
\newcommand{\mQ}{\textbf{\textit{Q}}}
\newcommand{\mR}{\textbf{\textit{R}}}
\newcommand{\mH}{\textbf{\textit{H}}} 

\newcommand{\mA}{\textbf{\textit{A}}}

\newcommand{\mI}{\textbf{\textit{I}}}

\newcommand{\mB}{\textbf{\textit{B}}}
\newcommand{\mC}{\textbf{\textit{C}}}
\newcommand{\mD}{\textbf{\textit{D}}}

\newcommand{\mU}{\textbf{\textit{U}}}
\newcommand{\mV}{\textbf{\textit{V}}}
\newcommand{\mE}{\textbf{\textit{E}}}
\newcommand{\mHr}{\textbf{\textit{H}}_r} 
\newcommand{\mHt}{\textbf{\textit{H}}_t} 
\newcommand{\mIr}{\textbf{\textit{I}}_{N_r}}

\newcommand{\setC}{\mathbb{C}} 

\newcommand{\vx}{\textbf{\textit{x}}}
\newcommand{\vy}{\textbf{\textit{y}}}
\newcommand{\vr}{\textbf{\textit{r}}}

\newcommand{\vn}{\textbf{\textit{n}}}

\newcommand{\va}{\textbf{\textit{a}}}

\newcommand{\vt}{\textbf{\textit{t}}}

\newcommand{\bPhi}{\boldsymbol{\Phi}}
\newcommand{\bUpsilon}{\boldsymbol{\Upsilon}}
\newcommand{\bTheta}{\boldsymbol{\Theta}}
\newcommand{\bPsi}{\boldsymbol{\Psi}}

\newcommand{\Pa}{P_{\mathrm{a}}} 
\newcommand{\Pamax}{P_{\mathrm{a}}^{\mathrm{max}}}
\newcommand{\Pbs}{P_{\mathrm{BS}}}
\newcommand{\an}{\alpha_n}
\newcommand{\ans}{\alpha_n^{\star}}

\begin{document}	
	\title{Hybrid Relay-Reflecting Intelligent Surface-Assisted Wireless Communication}
	\author{Nhan Thanh Nguyen, Quang-Doanh Vu, Kyungchun Lee, \IEEEmembership{Senior Member, IEEE}, and Markku Juntti, \IEEEmembership{Fellow, IEEE}
		\thanks{The research has been supported in part by Academy of Finland under 6Genesis Flagship (grant 318927) and EERA Project (grant 332362).}
		\thanks{N. T. Nguyen and Markku Juntti are with Centre for Wireless Communications, University of Oulu, P.O.Box 4500, FI-90014, Finland, (e-mail: nhan.nguyen@oulu.fi, markku.juntti@oulu.fi)}
		\thanks{Quang-Doanh Vu was with the Centre for Wireless Communications, University of
			Oulu. He is now with the Mobile Networks, Nokia,
			90650 Oulu, Finland (e-mail: quang-doanh.vu@nokia.com)}
		\thanks{K. Lee are with the Department of Electrical and Information Engineering, Seoul National University of Science and Technology, Seoul 01811, Republic of Korea (e-mail: kclee@seoultech.ac.kr).}
	}
	\maketitle	
	\begin{abstract}
		Reconfigurable intelligent surface (RIS) has emerged as a cost- and energy-efficient solution to enhance the wireless communication capacity. However, recent studies show that a very large surface is required for a RIS-assisted communication system; otherwise, they may be outperformed by the conventional relay. Furthermore, the performance gain of a RIS can be considerably degraded by hardware impairments such as limited-resolution phase shifters. To overcome those challenges, we propose a novel concept of hybrid relay-reflecting intelligent surface (HR-RIS), in which a single or few elements are deployed with power amplifiers (PAs) to serve as active relays, while the remaining elements only reflect the incident signals. Two architectures are proposed, including the fixed and dynamic HR-RIS. Their coefficient matrices are obtained based on alternating optimization (AO) and power allocation strategies, which enable understanding the fundamental performances of RIS and relaying-based systems with a trade-off between the two. The simulation results show that a significant improvement in both the spectral efficiency (SE) and energy efficiency (EE) with respect to the conventional RIS-aided system can be attained by the proposed schemes, especially, by the dynamic HR-RIS. In particular, the favorable design and deployment of the HR-RIS are analytically derived and numerically justified.
	\end{abstract}
	
	\IEEEpeerreviewmaketitle

	%========================= Introduction ========================================================
	\section{Introduction}
	
	The requirement of enhanced wireless communication services has created the demand to use higher frequency bands including the millimeter-wave (mmWave) bands. The high free-space path loss in those bands can be compensated for by using large antenna arrays at both ends of the link~\cite{Alkhateeb2014, He2014, Heath2016}. Therefore, massive multiple-input multiple-output (MIMO) transmission is built-in into the mmWave system design to attain a significant improvement in spectral efficiency (SE) from the large bandwidth and high beamforming gains. However, the mmWave massive MIMO system generally requires increased power consumption, hardware cost, and computational complexity for large-array signal processing  \cite{zhang2020capacity}. Furthermore, most practical mmWave links require line-of-sight (LoS) propagation, because otherwise, the received signal power is too low for reliable communications. Recently, a new technology called \emph{reconfigurable intelligent surface} (RIS), also known as intelligent reflecting surface (IRS)~\cite{Wu2019} or large intelligent surface (LIS)~\cite{Hu2018_SP} has emerged as a new promising solution to overcome the challenge of mmWave massive MIMO systems \cite{Liaskos18,2019Basar,Huang2018,di_renzo_smart_2019,He2019large}. It may enable LoS-type connectivity on the mmWave bands by steering a narrow beam from the RIS to the receiver assuming that LoS links exist between the transmitter and RIS as well as between the RIS and receiver. However, the potential benefits of a RIS do not limit to the mmWave bands but are of interest also for the sub-6~GHz frequencies.
	
	\subsection{Related Works}
	RIS is often assumed to be designed as a planar meta-surface or realized as a planar array of numerous \emph{passive reflecting elements}, which are connected to a controller, allowing modifying the phases of the incident signals independently in real-time. As a result, its phase shifts can be optimized to make the wireless channel between the transmitter and receiver more favorable for the communication \cite{zhang2020capacity, Wu2019, 2019Basar, Huang2018}. Several potential technologies for RIS implementation exist. It can be constructed as an array of discrete phase shifters, which control the impedance values of the elements. Thereby the phase shifts can steer the reflected signal to the desired direction of the receiver. Such RIS is called the \emph{discrete RIS} with no baseband processing capability~\cite{2019Basar,Huang2018,He2019large}. The phase shift control is typically assumed to be designed in some active network element such as a base station (BS) or access point (AP). The phase shift values are then transferred via some control link from the BS/AP to the RIS. Furthermore, a RIS with active receive elements introduced by Taha \textit{et al.} in~\cite{taha2019enabling, taha2019deep} has also been considered to enable the channel estimation at the RIS.
	
	The initial studies on RIS in the literature mainly focus on the performance and design aspects of the RIS-assisted communication systems. Specifically, Hu \textit{et. al} in \cite{hu2017potential,hu2018beyond} show that the capacity per RIS area-unit converges to $\frac{\bar{P}}{2\sigma^2}$ as the wavelength goes to zero, where $\bar{P}$ is the transmit power per area-unit, and $\sigma^2$ is the additive white Gaussian noise (AWGN) power spectral density. In \cite{he2020adaptive}, an adaptive phase shifter design based on hierarchical codebooks and limited feedback from the mobile station (MS) is proposed for a RIS-aided mmWave MIMO system for both accurate positioning and high data rate transmission. In \cite{jung2020asymptotic}, the asymptotic achievable rate in a RIS-assisted downlink system is analyzed under practical reflection coefficients, and a passive beamformer and a modulation scheme that can be used in a RIS without interfering with existing users is proposed to increase the achievable system sum-rate. Furthermore, in \cite{ozdogan2020using}, the propagation channel of point-to-point MIMO systems is enriched by using the RIS so that more paths with different spatial angles are added. As a result, the rank of the channel matrix is enhanced, and the multiplexing gain is achieved even when the direct path has low rank. The gain of RIS is further investigated in \cite{wu2019intelligent} and \cite{wang2020intelligent} for multiple-input-single-output (MISO) downlink channels operated at sub-6 GHz and mmWave frequencies, respectively. It is shown that the RIS of $N$ elements can achieve a total beamforming gain of $N^2$ \cite{wu2019intelligent} and allows the received signal power to increase quadratically with $N$ \cite{wang2020intelligent}. Moreover, the achievable data rate of communication systems assisted by a practical RIS with limited phase shifts and hardware impairments are characterized in \cite{zhang2020reconfigurable, guo2019weighted, han2019large, yang2020intelligentconf, zhang2020sum, alegria2019achievable}.
	
	Another important line of studies on RIS focuses on the capacity/data rate maximization of various RIS-assisted communication systems \cite{gong2020towards}. Specifically, in \cite{yang2020intelligent}, the problem of maximizing the achievable rate of a RIS-enhanced single-input-single-output (SISO)  orthogonal frequency division multiplexing (OFDM) system has been solved by jointly optimizing the transmit power allocation and the RIS passive array reflection coefficients. In contrast, the MISO systems have been considered in \cite{yu2019miso, yang2019irs, yuan2020intelligent, han2019large, di2020practical}. In particular, while \cite{yu2019miso, yang2019irs, yuan2020intelligent} focus on the joint transmit and passive beamforming problem of RISs with continuous phase shifts to attain substantially increased capacity/data rate, Han {\it et al.} in \cite{han2019large} consider the discrete phase shifts and justify that only a two-bit quantizer is sufficient to guarantee a high capacity, which agrees with the finding in \cite{di2020practical}. The capacity/rate optimization of RIS-assisted MIMO systems has been recently considered in \cite{zhang2020capacity, ying2020gmd, zhang2020sum}. In \cite{zhang2020capacity}, an efficient alternating optimization (AO) method has been proposed to find the locally optimal phase shifts of the RIS for both the narrowband frequency-flat and broadband frequency-selective MIMO OFDM systems. In particular, it is shown that with the proposed AO-based RIS design, the channel total power, rank, and condition number can be significantly improved for capacity enhancement. In \cite{zhang2020sum}, a RIS-enhanced full-duplex (FD) MIMO two-way communication system is considered. The system sum-rate is maximized through jointly optimizing the transmit beamforming and the RIS's coefficient matrix. Specifically, the non-convex optimization problem is decoupled into three sub-problems, which are solved iteratively, and its efficiency is justified by simulation results.
	
	\subsection{Motivations and Contributions}
	In this paper, we propose a novel \emph{semi-active RIS-aided beamforming} concept in which active beamforming (relaying) can be applied together with passive beamforming (reflecting). The idea is to activate a few elements of the RIS by connecting them with radio frequency (RF) chains and power amplifiers (PAs), allowing them to modify not only the phases but also the amplitudes of the incident signals. In this respect, these few activated elements become active amplify-and-forward (AF) relaying elements, and the conventional RIS becomes a \emph{hybrid relay-reflecting intelligent surface} (HR-RIS). We use this term throughout the paper to refer to the proposed architecture. 
	
	The HR-RIS may potentially be realized by recently introduced low-power complementary metal-oxide semiconductor (CMOS)-based technologies, such as the reflection amplifier (RA) \cite{landsberg2017design, landsberg2017low}, which can turn not only the phases but also the amplitudes of the incident signals. Thereby a possible implementation of the proposed architecture could produce a meta-surface with RA elements. However, this method can be cost-inefficient because the number of elements in the HR-RIS is large, while only a few active elements are required, as will be proven in the sequel. Therefore, a more practical solution appears to be upgrading the conventional RIS by replacing some elements with the RAs. Obviously, compared to the RIS, HR-RIS requires additional complexity for hardware implementation and signal processing of the active elements. However, those are only required for a single or few active elements. Considering that the total number of elements in the conventional RIS is very large, the hardware and computational costs increase only moderately. However, the detailed HR-RIS implementation is out of the scope of this paper. We focus on proposing it and analyzing its potential for system performance improvement.
	
	The HR-RIS is motivated by recent comparisons between the conventional relay and RIS, the practical deployment of RIS with hardware impairments, and the demand for channel estimation at the RIS. First, RIS can be thought in some respect similar to an FD relay \cite{bjornson_intelligent_2019, di2020reconfigurable}, but it consists of passive elements without the active power amplifier, enabling passive beamforming without requiring any active power and introducing no thermal noise. However, a main limitation of the RIS compared to the relays is the fact that the reflection limits the degrees of freedom in the beamforming. Therefore, HR-RIS, a hybrid architecture between the RIS and relay, can leverage the advantages while mitigating the disadvantages of RIS and relay. Second, based on the performance comparisons of relay and RIS in \cite{wu2019intelligent} and \cite{bjornson_intelligent_2019}, we found that the HR-RIS is promising to provide a remarkable performance improvement. Specifically, both Bj\"ornson {\it et al.} \cite{bjornson_intelligent_2019} and Wu {\it et al.} \cite{wu2019intelligent} show that a very large RIS is needed to outperform decode-and-forward (DF) relaying; otherwise, it can be easily outperformed even by a half-duplex (HD) relay with few elements. Furthermore, as the RIS becomes sufficiently large, the performance improvement is not significant if the number of elements in the RIS increases. This implies that if a few passive elements of the RIS are replaced by active ones, the reduction in the passive beamforming gain is just marginal, while the gain from active relaying can be significant. Third, it is shown in \cite{guo2019weighted, han2019large, yang2020intelligentconf, zhang2020sum} that RISs with limited-resolution phase shifts have considerable performance loss with respect to those with sufficiently large resolution. However, as the amplitudes of a few elements are adjustable, they can be optimized to compensate for the performance loss due to the limited-resolution phase shifts. The aforementioned aspects motivate a new architecture of the intelligent meta-surface, allowing it to leverage the benefits of both the RIS and relay, as the HR-RIS proposed in this paper.
	
	We recognize that there are several practical open problems related to the practical and efficient implementation of HR-RIS. What is more, assume perfect channel state information (CSI) \cite{wu2019intelligent, zhang2020capacity, guo2019weighted, pan2020multicell}. However, with the proposed HR-RIS architecture, the assumption is  less restrictive than with a passive RIS, because the active processing chains can be readily used for channel estimation using their built-in active elements \cite{taha2019enabling}. Thereby, our paper proposes the concept and demonstrates its significant system-level potential. This gives more understanding on the fundamental trade-offs between RIS and relaying as part of a MIMO communication link.
	
	The major contributions of the paper are summarized as follows:
	\begin{itemize}
		\item We propose the novel HR-RIS concept, which enables a hybrid active-passive beamforming scheme rather than fully-passive beamforming of the conventional RIS. It requires activating only a single or few elements of the RIS to serve as active relays to attain remarkable performance improvement. In particular, HR-RIS exploits the benefits of relaying while mitigating the limitation of passive reflecting.
		
		\item We propose two HR-RIS architectures, namely, \textit{fixed} and \textit{dynamic} HR-RIS. In the former, the active elements are fixed in manufacture; by contrast, those in the latter can be adaptively configured to improve the performance and save power consumption. The coefficient matrices of the HR-RIS are optimized in the formulated SE maximization problem, which is solved by the AO and power allocation schemes. Furthermore, the favorable deployment and performance gains of the HR-RIS are derived analytically and numerically verified by computer simulations.
		
		\item The total power consumption and energy efficiency of the system with the proposed HR-RIS is analyzed and compared to that with the conventional RIS. It is shown that the former requires higher power consumption than the latter because additional power is consumed for the active processing. However, with a small number of active elements in the proposed HR-RIS, the improvement in both the SE and energy efficiency (EE) is guaranteed, as numerically demonstrated by the simulation results.
		
		\item The paper bridges the theoretical gap between a passive RIS and active AF relay aiding a MIMO communication link. This gives more insight into the fundamental performance of each of the approaches and attainable SE and EE.
	\end{itemize}
	
	\textit{Structure:} The rest of the paper are as follows. In Section \ref{sec_system_model}, we introduce the concept of the HR-RIS and the system model of the HR-RIS-aided MIMO system, and the problem of SE maximization is formulated. Its efficient solution is found in Section \ref{sec_efficient_solution}. Based on that, the coefficient matrices of the fixed and dynamic HR-RIS architectures are derived in Section \ref{sec_HRRIS_architecture}. Their power consumption is investigated in Section \ref{sec_power}. Simulation results are shown in Section \ref{sec_sim}, and, finally, conclusions are drawn in Section \ref{sec_concusion}. 
	
	\textit{Notations}: Throughout this paper, numbers, vectors, and matrices are denoted by lower-case, bold-face lower-case, and bold-face upper-case letters, respectively. $(\cdot)^*$ and $(\cdot)^H$ denote the conjugate of a complex number and the conjugate transpose of a matrix or vector, respectively. $\mI_N$ denotes the identity matrix of size $N \times N$, and $\mathrm {diag} \{ a_1, \ldots, a_N \}$ represents a diagonal matrix with diagonal entries $a_1, \ldots, a_N$. Furthermore, $\abs{\cdot}$ denotes either the absolute value of a scalar or determinant of a matrix, and the expectation operator is denoted by $\mean{\cdot}$.

	%========================= System model and Problem formulation ===============================
	\section{HR-RIS Architecture, System model, and SE Problem formulation}
	\label{sec_system_model}
	
	\begin{figure*}[t]
		%\vspace{-0.5cm}
		\belowcaptionskip = -0.5cm
		\centering
		\includegraphics[scale=0.55]{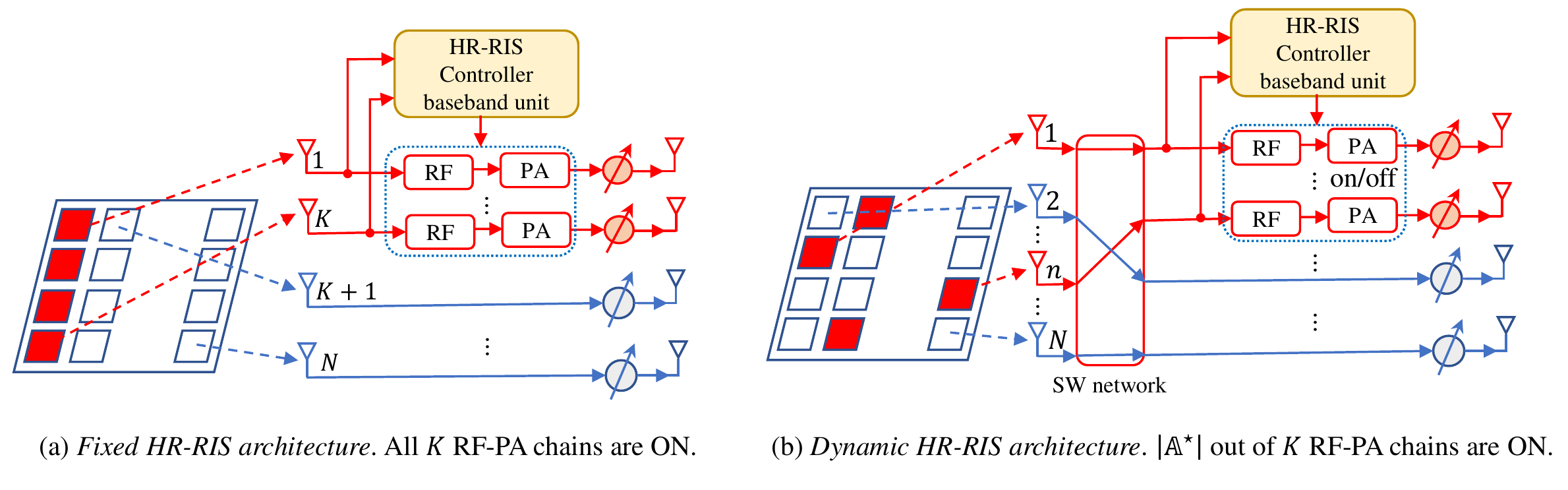}
		\caption{Illustration of the fixed and dynamic HR-RIS architectures.}
		\label{fig_HRRIS_architecture}
	\end{figure*}
	
	\subsection{Proposed HR-RIS Beamforming Architecture}
	
	The HR-RIS is equipped with $N$ elements, including $M$ passive reflecting and $K$ active relay elements $(M+K=N)$. Here, the passive element can only shift the phase while the active one can tune both the phase and amplitude of the incident signal. As such, for $K=0$, the HR-RIS serves as the conventional RIS; by contrast, for $K=N$, it becomes a conventional relay station. It is noted that an active element would consume more power for processing compared to the passive one due to the operation of the RF chain. Thus, for practical deployment, we are interested in the case that $1 \leq K  \ll M$. Furthermore, similar to the conventional RIS, we assume that each (active/passive) element of the HR-RIS can re-scatter the signal independently (without cross-interference) by an individual coefficient \cite{zhang2020capacity}.
	
	We now introduce some notations in order to mathematically model the HR-RIS. In particular, let us denote by $\setA$, $\setA \subset \{1,2,\ldots,N\}$, the index set of the positions of the $K$ active elements. Let $\an$ denote the relay/reflection coefficient used at the $n$th element, which is given as 
	\begin{align}
		\label{def_alpha}
		\an =
		\begin{cases}
			\abs{\an} e^{j \theta_n}, & \text{if } n \in \setA \\
			e^{j \theta_n}, & \text{otherwise}
		\end{cases},
	\end{align}
	where $\theta_n \in [0, 2\pi)$ represent the phase shift. We note that $\abs{\an} = 1$ for $n \notin \setA$. We also define three diagonal matrices constructed from $\{\alpha_n\}_{n=1,\ldots,N}$. Let  $\bUpsilon = \text{diag} \{ \alpha_1, \ldots, \alpha_{N} \} \in \setC^{N \times N}$ be the diagonal matrix of the coefficients. We define an additive decomposition for it as $\bUpsilon = \bPhi + \bPsi$, where
	$\bPhi = \text{diag} \{ \phi_1, \ldots, \phi_{N} \} \in \setC^{N \times N}$, $\bPsi = \text{diag} \{ \psi_1, \ldots, \psi_{N} \} \in \setC^{N \times N}$, 
	\begin{align*}
		\phi_n &=
		\begin{cases}
			0, & \text{if } n \in \setA \\
			\an = e^{j \theta_n}, & \text{otherwise}
		\end{cases}, \\
		\psi_n &=
		\begin{cases}
			\an = \abs{\an} e^{j \theta_n}, & \text{if } n \in \setA \\
			0, & \text{otherwise}
		\end{cases}.
	\end{align*} 
	In other words, $\bPhi$ and $\bPsi$ contain the passive reflecting and active relaying coefficients, respectively.%, while $\bUpsilon$ contains the coefficients of all the elements of HR-RIS. Clearly, we have $\bUpsilon = \bPhi + \bPsi$.
	
	There can be two architectures for the HR-RIS, namely, the fixed and dynamic HR-RIS. In the former, the number and positions of the active elements, i.e., set ${\setA}$, are predefined and fixed as illustrated in Fig.\ \ref{fig_HRRIS_architecture}(a) \cite{taha2019deep, taha2019enabling}. By contrast, in the dynamic HR-RIS architecture, the number and positions of active elements can be dynamically changed according to the propagation condition. In other words, set ${\setA}$ can be a design parameter in this architecture. More specifically, there are a number of RF-PA chains each can be turned on/off. An element of HR-RIS is active if it connects to an  ``ON" RF-PA chain; otherwise, it serves as a passive reflecting element \cite{taha2019deep, taha2019enabling, wu2020intelligent}. Such connection can be done via a switching (SW) network as illustrated in Fig.\ \ref{fig_HRRIS_architecture}(b). Furthermore, the signal processing in the active elements are controlled by an HR-RIS controller and baseband unit \cite{taha2019deep, taha2019enabling}. In this paper, both the fixed and dynamic HR-RIS architectures are considered.
	
	\subsection{System Model}

	\begin{figure}[t]
		\belowcaptionskip = -0.5cm
		\includegraphics[scale=0.5]{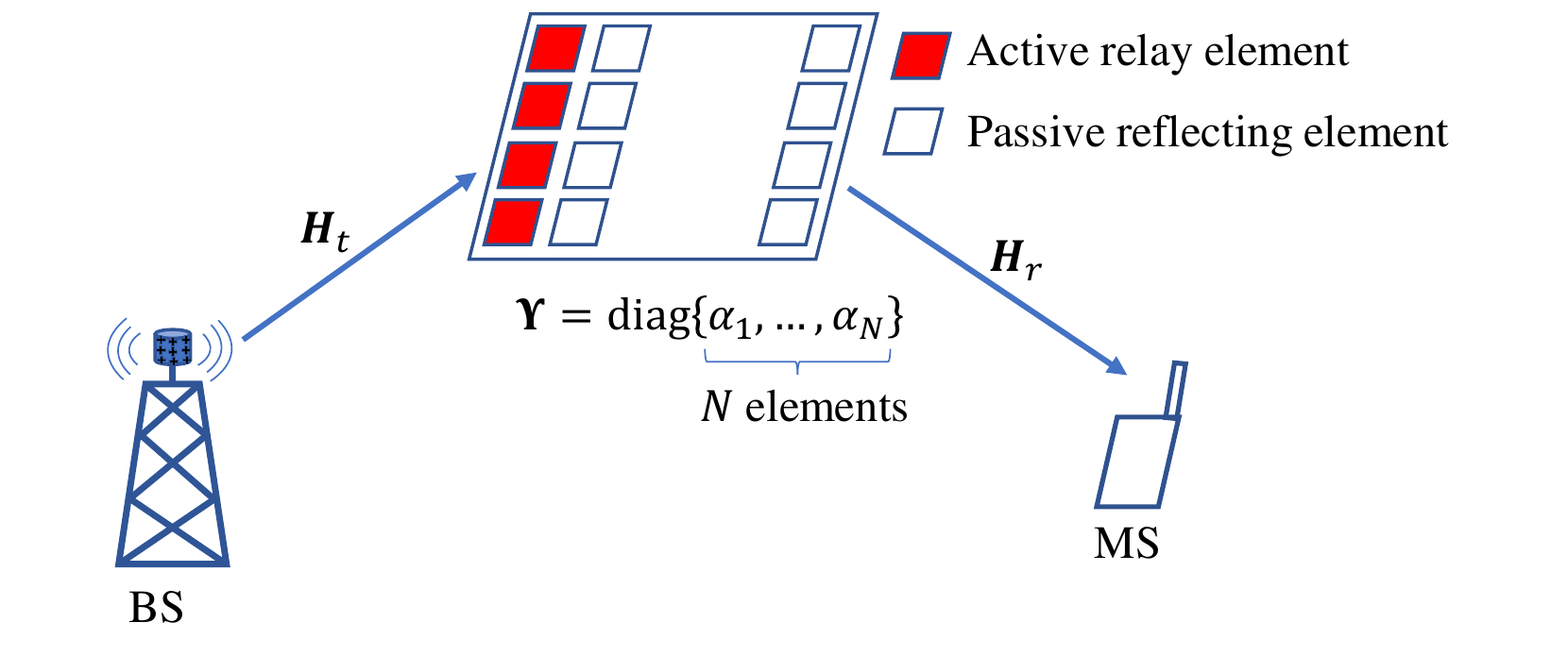}
		\caption{Illustration of the HR-RIS-aided communication system.}
		\label{Fig_HRRIS_system}
	\end{figure}
	
	We consider a downlink transmission between a BS and an MS. Suppose that there is no direct link between the BS and the MS due to, e.g., severe pathloss or blockage \cite{han2020cooperative}, and the communications is aided by the HR-RIS as illustrated in Fig.\ \ref{Fig_HRRIS_system} \cite{taha2019deep,taha2019enabling}. Let us denote by $N_t$ and $N_r$ the numbers of antennas equipped at the BS and MS, respectively, where $N_t\geq 1$ and  $N_r \geq 1$. Let $\mHt \in \setC^{N \times N_t}$ and $\mHr \in \setC^{N_r \times N}$ denote the channel between the BS and the HR-RIS, and between the HR-RIS and the MS, respectively. Let $\vx \in \setC^{N_t \times 1}$ be the transmitted signal vector, where $\mean{\vx \vx^H} = \Pbs \mI_{N_t}$ and $\Pbs$ is the transmit power at the BS. With these newly introduced notations, the received signals at the MS can be written as
	\begin{align*}
		\vy &= \mHr \bPhi \mHt \vx + \mHr \bPsi \mHt \vx + \mHr \bPsi \vn_{\mathrm{H}} + \vn_{\mathrm{MS}} \\
		&= \left(\mHr \bPhi \mHt  + \mHr \bPsi \mHt\right) \vx + \vn \\
		&= \mHr \bUpsilon \mHt \vx  + \vn, \nbthis \label{eq_system_RIS}
	\end{align*}
	where $\vn_{\mathrm{H}} \sim \mathcal{CN} (0, \sigma_{\mathrm{H}}^2 \mI_{K})$ and $\vn_{\mathrm{MS}} \sim \mathcal{CN} (0, \sigma_{\mathrm{MS}}^2 \mIr)$ are the complex additive white Gaussian noise (AWGN) vectors at the $K$ active relay elements of the HR-RIS and at the MS, respectively; and $\vn = \mHr \bPsi \vn_{\mathrm{H}} + \vn_{\mathrm{MS}}$ represents the total effective noise at the MS. For notational simplicity, we assume that  $\sigma_{\mathrm{H}}^2 = \sigma_{\mathrm{MS}}^2 = \sigma^2$ and $\vn \sim \mathcal{CN} \left(0, \sigma^2\left(\mIr + \mHr \bPsi \bPsi^H \mHr^H\right)\right)$.

	\subsection{Problem Formulation}
	
	Based on \eqref{eq_system_RIS}, the SE of an HR-RIS-aided MIMO system can be expressed as
	\begin{align*}
		f_0(\{\alpha_n\}) =  \log_2 \abs{ \mIr + \rho \mHr \bUpsilon \mHt \mHt^H \bUpsilon^H \mHr^H \mR^{-1}},
	\end{align*}
	where $\{\alpha_n\}$ represents the set of all the relay/reflecting coefficients on the main diagonal of $\bUpsilon$, i.e., $\{\alpha_n\} = \{\alpha_1, \alpha_2, \ldots, \alpha_N\}$. Furthermore,
	\begin{align*}
		\mR = \mIr + \mHr \bPsi \bPsi^H \mHr^H  \in \setC^{N_r \times N_r} \nbthis \label{eq_Rn}
	\end{align*}
	is the aggregate noise covariance matrix, and $\rho = \frac{\Pbs}{\sigma^2}$.
	The transmit power of the active elements of the HR-RIS is  
	\begin{align*}
		\Pa(\{\alpha_n\}) &\triangleq \tr{ \bPsi \left( \mHt \mean{\vx \vx^H} \mHt^H + \sigma^2 \mI_{N} \right) \bPsi^H}  \\
		&= \tr{ \bPsi \left( \Pbs \mHt \mHt^H + \sigma^2 \mI_{N} \right) \bPsi^H}. \nbthis \label{eq_Pactive}
	\end{align*}
	
	The problem of designing relay/reflection coefficients of the HR-RIS for maximizing the SE can be formulated as
	\begin{subequations}
		\begin{align}
			(\mathrm{P0}) \quad \underset{\{\alpha_n\}}{\textrm{maximize}} \quad &  f_0(\{\alpha_n\}) \label{opt_hyb_obj_SE} \\
			\textrm{subject to} \quad 
			& \abs{\alpha_n}=1 \text{ for } n \notin \setA \label{opt_hyb_con_unit_modul}\\
			&\Pa(\{\alpha_n\}) \leq \Pa^{\textrm{max}} \label{opt_hyb_con_relay_power}
		\end{align}
	\end{subequations}
	where $\Pa^{\textrm{max}}$ is the  power budget of the active elements. Function $f_0(\{\alpha_n\})$ is nonconvex with respect to $\{\alpha_n\}$. In addition, the feasible set of $(\mathrm{P0})$ is nonconvex due to the unit-modulus constraint \eqref{opt_hyb_con_unit_modul}. Consequently, problem $(\mathrm{P0})$ is intractable, and it is difficult to find an optimal solution. In the following section, we develop an efficient solution to $(\mathrm{P0})$.
	
	\section{Efficient Solution to  $(\mathrm{P0})$}
	\label{sec_efficient_solution}
	\subsection{A Tractable Approximation of $(\mathrm{P0})$}
	
	As the first step of developing an efficient solution to $(\mathrm{P0})$, we approximate the problem into a more tractable one. Specifically, $f_0(\{\alpha_n\})$ is upper bounded by $f(\{\alpha_n\})$, as follows:
	\begin{align*}
		f_0(\{\alpha_n\})
		&=  \log_2 \abs{ \mIr + \rho \mHr \bUpsilon \mHt \mHt^H \bUpsilon^H \mHr^H \mR^{-1}} \\
		& = \log_2 \abs{ \mR + \rho \mHr \bUpsilon \mHt \mHt^H \bUpsilon^H \mHr^H } - \log_2 \abs{\mR}  \nbthis \label{eq_ub_2} \\
		&\leq \log_2 \abs{ \mIr + \mHr \bPsi \bPsi^H \mHr^H  + \rho \mHr \bUpsilon \mHt \mHt^H \bUpsilon^H \mHr^H }  \nbthis \label{eq_ub_4} \\
		&\triangleq f(\{\alpha_n\}),
	\end{align*}
	where the inequality in \eqref{eq_ub_4} follows by substituting (\ref{eq_Rn}) to (\ref{eq_ub_2}). Clearly, the equality occurs when $\setA = \emptyset$. In addition, the smaller the term $\log_2 \abs{\mR}$ is, the tighter the bound is. We can also observe that $\log_2 \abs{\mR}$ is small when the path loss is large and the number of active elements is small (since $\bPsi$ becomes very sparse when only a single or few active elements are employed at the HR-RIS). These observations motivate us to arrive at the following approximation of $(\mathrm{P0})$
	\begin{align*}
		\phybrid \quad \underset{\{\alpha_n\}}{\textrm{maximize}} \quad &  f(\{\alpha_n\}) \\
		\textrm{subject to} \quad &\eqref{opt_hyb_con_unit_modul},\eqref{opt_hyb_con_relay_power}.
	\end{align*}
	Although $f(\{\alpha_n\})$ is still nonconvex with respect to $\{\alpha_n\}$, its structure allows us to develop an efficient solution via the AO method. The details are presented in the following subsection.
	
	\subsection{Efficient Solution to $\phybrid$: an AO Approach}\label{sec_sol_fixHR}
	
	Generally, the proposed solution is a sequential procedure: in each iteration, a specific coefficient of HR-RIS is updated when the others are fixed. We first further expand $f(\{\alpha_n\})$ to extract the role of variable $\alpha_n$, for any $n \in \{1,...,N\}$. Specifically, let $\vr_n \in \setC^{N_r \times 1}$ be the $n$th column of $\mHr$, and $\vt_n^H \in \setC^{1 \times N_t}$ be the $n$th row of $\mHt$, i.e., $\mHr = [\vr_1, \ldots, \vr_N]$ and $\mHt = [\vt_1, \ldots, \vt_N]^H$. Since $\bUpsilon$ and $\bPsi$ are diagonal matrices, we have $\mHr \bUpsilon \mHt = \sum_{n=1}^{N} \an \vr_n \vt_n^H$ and $\mHr \bPsi = \sum_{n \in \setA} \an \vr_n$. Consequently, we can express $f(\{\alpha_n\})$ as follows
	\begin{align}
		f(\{\alpha_n\}) = \log_2 \abs{\mA_n +  \abs{\an}^2 \mB_n + \an \mC_n + \an^* \mC_n^H}, \label{fm}
	\end{align}
	%which leads to
	%\begin{align*}
	%&\mIr + (\mHr \bPsi) (\mHr \bPsi)^H +  \rho(\mHr \bUpsilon \mHt) (\mHr \bUpsilon \mHt)^H  \\
	%&= \mIr + \left(\sum_{n \in \setA} \an \vr_n \right) \left(\sum_{n\in \setA} \an \vr_n \right)^H \\
	%&\hspace{2.5cm} + \rho\left(\sum_{n=1}^{N} \an \vr_n \vt_n^H\right) \left(\sum_{n=1}^{N} \an \vr_n \vt_n^H\right)^H  \\
	%&= \mA_n +  \abs{\an}^2 \mB_n + \an \mC_n + \an^* \mC_n^H, %\nbthis \label{eq_ABC}
	%\end{align*}
	where matrices $\mA_n$, $\mB_n$, and $\mC_n$ are given by
	\begin{align*}
		\mA_n &\triangleq \mIr + \left( \sum_{i \in \setA, i \neq n} \alpha_i \vr_i \right) \left( \sum_{i\in \setA,i \neq n} \alpha_i^* \vr_i^H\right) \\
		&\quad + \rho \left( \sum_{i=1, i \neq n}^{N} \alpha_i \vr_i \vt_i^H\right) \left( \sum_{i=1,i \neq n}^{N} \alpha_i^* \vt_i \vr_i^H\right), \nbthis \label{def_A_active} \\
		\mB_n &\triangleq \vr_n \vr_n^H + \rho \vr_n \vt_n^H \vt_n \vr_n^H , \nbthis \label{def_B_active} \\
		\mC_n &\triangleq \vr_n  \left( \sum_{i\in \setA, i \neq n} \alpha_i^* \vr_i^H \right) + \rho \vr_n  \vt_n^H \left( \sum_{i=1, i \neq n}^{N} \alpha_i^*  \vt_i \vr_i^H\right)  \nbthis \label{def_C_active} 
	\end{align*}
	for $n \in \setA$, and
	\begin{align*}
		\mA_n &\triangleq \mIr + \rho \left( \sum_{i=1, i \neq n}^{N} \alpha_i \vr_i \vt_i^H\right) \left( \sum_{i=1,i \neq n}^{N} \alpha_i^* \vt_i \vr_i^H\right), \nbthis \label{def_A_passive}  \\
		\mB_n &\triangleq \rho \vr_n \vt_n^H \vt_n \vr_n^H ,  \nbthis \label{def_B_passive}\\
		\mC_n &\triangleq \rho \vr_n \vt_n^H \left( \sum_{i=1, i \neq n}^{N} \alpha_i^* \vt_i \vr_i^H\right) \nbthis \label{def_C_passive}
	\end{align*}
	for $n \notin \setA$. We note that matrices $\mA_n, \mB_n,$ and $\mC_n$ do not contain variable $\alpha_n$. This means that if all variables $\{\alpha_i\}_{i=1,i \neq n}^{N}$ are fixed, the three matrices are determined.
	
	Similarly, we also extract the role of variable $\alpha_n$ in $\Pa(\{\alpha_n\})$ by rewriting
	\begin{align*}
		\Pa(\{\alpha_n\}) &= \tr{ \Pbs \bPsi \mHt \mHt^H \bPsi^H + \sigma^2 \bPsi  \bPsi^H}  \\
		&= \sigma^2 \tr{\bPsi  \bPsi^H} + \Pbs \tr{\bPsi \mHt \mHt^H \bPsi^H}\\
		&= \sigma^2 \sum_{n \in \setA} \abs{\an}^2 + \Pbs \sum_{n \in \setA} \abs{\an}^2 \norm{\vt_n}^2  \\
		&= \sum_{n \in \setA} \abs{\an}^2 \xi_n, \nbthis \label{eq_relay_power_sum}
	\end{align*}
	where $\xi_n \triangleq \sigma^2 + \Pbs\norm{\vt_n}^2$. Let $\tilde{\Pa}  \triangleq \sum_{i \in \setA, i \neq n} \abs{\alpha_i}^2 \xi_i$, $n\in\setA$. We can see that 	$\tilde{\Pa}$ is a constant if all variables $\{\alpha_i\}_{i=1,i \neq n}^{N}$ are fixed. Therefore, the role of $\alpha_n$ in $\Pa(\{\alpha_n\})$ can be seen in the following expression of $\Pa(\{\alpha_n\})$:
	\begin{align*}
		\Pa(\{\alpha_n\}) = \abs{\an}^2 \xi_n +\tilde{\Pa}. \nbthis \label{eq_relay_power}
	\end{align*}
	
	\subsubsection{Problem of Updating $\an$}
	In the AO approach, in an iteration of updating $\alpha_n$, $\{\alpha_i\}_{i=1,i \neq n}^{N}$ are fixed. Therefore, the objective function $f(\{\alpha_n\})$ in \eqref{fm} can be rewritten in the explicit form
	\begin{align}
		f_n(\alpha_n) = \log_2 \abs{\mA_n +  \abs{\an}^2 \mB_n + \an \mC_n + \an^* \mC_n^H}, \label{fn}
	\end{align}
	which has only one variable, i.e., $\alpha_n$. Moreover, because $\mA_n$ is of full-rank ($\textrm{rank}(\mA_n) = N_r$) and invertible, we can rewrite $f_n(\alpha_n)$ as  
	\begin{align*}
		&f_n(\alpha_n) = \log_2 \abs{\mA_n} + g_n(\alpha_n), \nbthis \label{def_fn}
	\end{align*} 
	where
	\begin{align*}
		&g_n(\alpha_n)  \triangleq \log_2 \abs{\mIr +  \abs{\an}^2 \mA_n^{-1}\mB_n + \an \mA_n^{-1}\mC_n + \an^* \mA_n^{-1}\mC_n^H}. \nbthis \label{def_gn}
	\end{align*}
	In \eqref{def_fn}, $\log_2 \abs{\mA_n}$ is a constant with $\alpha_n$. Therefore, the problem of updating $\alpha_n$, denoted by $\pupdate$, is given by 
	\begin{align}
		\begin{cases}
			\underset{\alpha_n}{\textrm{maximize}}\;g_n(\alpha_n)\;\textrm{subject to}\;\abs{\an} = 1, & \textrm{for}\;n \notin \setA \\
			\underset{\alpha_n}{\textrm{maximize}}\;g_n(\alpha_n)\;\textrm{subject to}\;\abs{\an}^2 \leq \frac{\Pamax - \tilde{\Pa} }{\xi_n}, & \textrm{for}\;n\in \setA
		\end{cases} \label{probupdate}
	\end{align}
	
	\subsubsection{Solution to $\pupdate$}
	$\pupdate$ in \eqref{probupdate} admits a closed-form solution, and, thus, it is efficient for practical implementation. To derive it, we rewrite $g_n(\alpha_n)$ as
	\begin{align*}
		g_n(\alpha_n) &= \log_2 \abs{\mD_n + \an \mA_n^{-1}\mC_n + \an^* \mA_n^{-1}\mC_n^H}\\
		&= \log_2 \abs{\mD_n} + \log_2 \abs{ \mIr + \an \mE_n^{-1}\mC_n + \an^* \mE_n^{-1}\mC_n^H}, \nbthis \label{def_h_n}
	\end{align*}
	where $\mD_n = \mIr + \abs{\an}^2 \mA_n^{-1}\mB_n$, and $\mE_n = \mA_n \mD_n$. We start investigating the objective function $g_n(\alpha_n)$ by considering the first term in \eqref{def_h_n}, i.e., $\log_2 \abs{\mD_n}$. Specifically, we note that in $\mD_n$, $\mathrm{rank} (\mA_n^{-1}\mB_n) \leq \mathrm{rank} (\mB_n) = 1$. In addition, the probability that $\mathrm{rank} (\mA_n^{-1}\mB_n) = 0$ is almost zero (it only happens when $\mA_n^{-1}\mB_n = \textbf{0}$). Consequently, we generally have $\mathrm{rank} (\mA_n^{-1}\mB_n) = 1$. Also, we observe that  $\mA_n^{-1}\mB_n$ is non-diagonalizable iff $\mathrm{trace} (\mA_n^{-1}\mB_n) = 0$ \cite{zhang2020capacity}, which rarely happens in general. Thus, we almost surely have $\mathrm{trace} (\mA_n^{-1}\mB_n) \neq 0$ and $\mA_n^{-1}\mB_n$ is diagonalizable. As a result, we can write as $\mA_n^{-1}\mB_n = \mQ_n \boldsymbol{\Gamma}_n \mQ_n^{-1}$ based on the eigenvalue decomposition (EVD) where $\mQ_n \in \setC^{N_r \times N_r}$, $\boldsymbol{\Gamma}_n = \diag{\gamma_n, 0, \ldots, 0}$, and $\gamma_n$ is the sole non-zero eigenvalue of $\mA_n^{-1}\mB_n$. Finally, since $\mA_n$ and $\mB_n$ are both positive semi-definite, $\gamma_n$ is non-negative and real. Thus, we can write
	\begin{align*}
		\log_2 \abs{\mD_n} &= \log_2 \abs{\mIr + \abs{\an}^2 \mQ_n \boldsymbol{\Gamma_n} \mQ_n^{-1}} \\
		&= \log_2 \abs{\mQ_n \left(\mIr + \abs{\an}^2  \boldsymbol{\Gamma_n}\right) \mQ_n^{-1}}\\
		&= \log_2 \abs{\mIr + \abs{\an}^2  \boldsymbol{\Gamma_n}}\\
		& = \log_2 \left(1 + \abs{\an}^2  \gamma_n \right). \nbthis \label{eq_det_Dn}
	\end{align*}

	We now focus on the second term in \eqref{def_h_n}. Following the similar arguments for the first term, we also almost surely have $\mE_n^{-1}\mC_n$ diagonalizable. Thus, we have $\mE_n^{-1}\mC_n = \mU_n \boldsymbol{\Lambda}_n \mU_n^{-1}$ based on the EVD, where $\mU_n \in \setC^{N_r \times N_r}$, $\boldsymbol{\Lambda}_n = \diag{\lambda_n, 0, \ldots, 0}$, and $\lambda_n$ is the sole non-zero eigenvalue of $\mE_n^{-1}\mC_n$. Now, let $\mV_n \triangleq \mU_n^H \mE_n \mU_n$, $v_n$ be the first element of the first column of $\mV_n^{-1}$, and $v_n^{\prime}$ be the first element of the first row of $\mV_n$. As a result, we can write \cite{zhang2020capacity}
	\begin{align*}
		&\log_2 \abs{ \mIr + \an \mE_n^{-1}\mC_n + \an^* \mE_n^{-1}\mC_n^H} \\
		&= \log_2 \left(1 + \abs{\an}^2 \abs{\lambda_n}^2  + 2\re{\an \lambda_n} - v_n^{\prime} v_n \abs{\lambda_n}^2 \right), \nbthis \label{eq_g_x}
	\end{align*}
	where $\re{\cdot}$ denotes the real part of a complex number. The detailed derivation of \eqref{eq_g_x} can be found in \cite{zhang2020capacity}. We note that, compared to \cite{zhang2020capacity}, the additional coefficient $\abs{\an}^2$ is due to the active relaying coefficient in the HR-RIS, which does not exist for the conventional RIS  \cite{zhang2020capacity}. 
	
	In summary, based on \eqref{eq_det_Dn} and \eqref{eq_g_x}, we have
	\begin{align*}
		&g_n(\alpha_n) =  \log_2 \left(1 + \abs{\an}^2  \gamma_n \right) \\
		& + \log_2 \left(1 + \abs{\an}^2 \abs{\lambda_n}^2  + 2\re{\an \lambda_n} - v_n^{\prime} v_n \abs{\lambda_n}^2 \right). \nbthis \label{eq_fn}
	\end{align*}
	It is observed that an optimal solution to $\pupdate$, denoted by $\ans$, admits the form
	\begin{align}
		\label{eq_alpha_best}
		\ans =
		\begin{cases}
			\abs{\ans} e^{-j \arg \{ \lambda_n \}}, & n \in \setA\\
			e^{-j \arg \{ \lambda_n \}}, & \text{otherwise}
		\end{cases}.
	\end{align}
	It is observed that the amplitude $\abs{\ans}$ is not available at this stage because it requires a determined $\setA$. To this end, it is natural to consider two scenarios: $\setA$ is predetermined and fixed in the manufacture, and $\setA$ is dynamic and optimized based on the propagation condition, associated with the fixed and dynamic HR-RIS architectures illustrated in Figs.\ \ref{fig_HRRIS_architecture}(a) and \ref{fig_HRRIS_architecture}(b), respectively. The solution $\{\ans\}$ dedicated to these architectures will be derived in the next section.
	
	\section{Beamforming Coefficient Matrix of Fixed and Dynamic HR-RIS}
	\label{sec_HRRIS_architecture}
	
	\subsection{Fixed HR-RIS Architecture}
	
	In the fixed HR-RIS, $\setA$ is available to determine $\{\abs{\ans}\}_{n \in \setA}$. Specifically, $\abs{\ans}$, $n \in \setA$, can be determined based on the fact that, given the optimal form in  \eqref{eq_alpha_best}, $g_n(\ans)$  monotonically increases with $\abs{\ans}$. Therefore, from \eqref{probupdate}, we obtain 
	\begin{align}
		\abs{\ans} = \sqrt{\frac{\Pamax - \tilde{\Pa} }{\xi_n}}, n \in \setA. \label{eq_optimal_active_amplitude}
	\end{align}
	As a result, the optimal solution to $\pupdate$ is given as
	\begin{align}
		\label{eq_alpha_best_fix}
		\ans =
		\begin{cases}
			\sqrt{\frac{\Pamax - \tilde{\Pa} }{\xi_n}} e^{-j \arg \{ \lambda_n \}}, & n \in \setA\\
			e^{-j \arg \{ \lambda_n \}}, & \text{otherwise}
		\end{cases}.
	\end{align}
	
	The solution \eqref{eq_alpha_best_fix} and the given $\setA$ are readily used to obtain the coefficient matrix of the fixed HR-RIS scheme, as outlined in Algorithm \ref{alg_fix}. Specifically, at the initial stage, coefficients $\{\an\}$ are randomly generated to have arbitrary phases and amplitudes satisfying $\abs{\an} = 1, n \notin \setA$ and $\sum_{n \in \setA} \abs{\an}^2 \xi_n = \Pamax$, based on \eqref{opt_hyb_con_relay_power}, \eqref{eq_relay_power_sum}, and \eqref{eq_optimal_active_amplitude}. During steps 2--13, the coefficients are alternatively updated based on \eqref{eq_alpha_best_fix}. It terminates when a convergence criteria on the objective $f(\{\alpha_n\})$ is reached. The iteration procedure in Algorithm \ref{alg_fix} guarantees the convergence. To see this, we note that objective function $f(\{\alpha_n\})$ is upper bounded. In addition, $\pupdate$ is solved optimally by \eqref{eq_alpha_best_fix}. Thus, the resultant sequence of $f(\{\alpha_n\})$ is non-decreasing over iterations \cite{zhang2020capacity}. We make two remarks on the design and deployment of the fixed HR-RIS.

	\begin{remark}
		\label{rm_gain_vs_K}
		Inserting $\tilde{\Pa}  \triangleq \sum_{i \in \setA, i \neq n} \abs{\alpha_i}^2 \xi_i$, $n\in\setA$ to \eqref{eq_optimal_active_amplitude}, it is observed that a larger $K$ results in a smaller $\abs{\ans}$. Therefore, increasing $K$ does not always guarantee the SE improvement of the fixed HR-RIS with respect to the conventional RIS. In particular, with a limited power budget $\Pamax$, the HR-RIS can have $\abs{\ans} < 1$, causing signal attenuation, and hence degrades the SE. In this case, the fixed HR-RIS with a small $K$ is easier to attain SE gains than that with numerous active elements. Furthermore, in the fixed HR-RIS, we have $\sum_{n \in \setA} \abs{\an}^2 \xi_n = \Pamax$, implying that $\Pamax > \sum_{n \in \setA} \xi_n$ is necessary for $\abs{\an} > 1, \forall n \in \setA$.
	\end{remark}

	\begin{remark}
		\label{rm_gain_Pt}
		By inserting $\xi_n \triangleq \sigma^2 + \Pbs\norm{\vt_n}^2$ to \eqref{eq_optimal_active_amplitude}, it is observed that with a fixed $\Pamax$, a lower transmit power $\Pbs$ and/or smaller channel gain $\norm{\vt_n}^2$ from the transmitter results in a higher power amplifier coefficients $\abs{\ans}$, and thus, a more significant performance improvement can be achieved.
	\end{remark}

	Remarks \ref{rm_gain_vs_K} and \ref{rm_gain_Pt} will be further numerically justified and discussed in Section \ref{sec_sim}. The fixed number and positions of active elements in the fixed HR-RIS limit the beamforming gain of the HR-RIS. In the next section, we propose a more robust architecture, named the dynamic HR-RIS, which overcomes the drawbacks of the fixed HR-RIS to guarantee the SE improvement.
	
	\begin{algorithm}[t]
		%\small
		\caption{Find coefficients of the fixed HR-RIS}
		\label{alg_fix}
		\begin{algorithmic}[1]
			\REQUIRE $\mHt, \mHr, \setA$.
			\ENSURE $\{\alpha_1^{\star}, \ldots, \alpha_N^{\star}\}$.
			\STATE Randomly generate $\{\alpha_n\}$ with $\abs{\an}=1$, $n \notin \setA$, and $\sum_{n \in \setA} \abs{\an}^2 \xi_n = \Pamax$.
			\WHILE{objective value does not converge }
			\FOR {$n = 1 \rightarrow N$}
			\IF {$n \in \setA$}
			\STATE Compute $\mA_n, \mB_n$, and $\mC_n$ based on \eqref{def_A_active}-\eqref{def_C_active}.
			\STATE $\mD_n = \mIr + \abs{\an}^2 \mA_n^{-1}\mB_n$, $\mE_n = \mA_n \mD_n$.
			\ELSE
			\STATE Compute $\mA_n, \mB_n$, and $\mC_n$ based on \eqref{def_A_passive}-\eqref{def_C_passive}.
			\STATE $\mD_n = \mIr + \mA_n^{-1}\mB_n$, $\mE_n = \mA_n \mD_n$.
			\ENDIF
			\STATE Find $\lambda_n$ as the sole non-zero eigenvalue of $\mE_n^{-1}\mC_n$.\\
			\STATE Update $\ans$ as \eqref{eq_alpha_best_fix}.
			\ENDFOR 
			\ENDWHILE
		\end{algorithmic}
	\end{algorithm}
	
	\subsection{Dynamic HR-RIS Architecture}
	Unlike the fixed HR-RIS architecture, in the dynamic HR-RIS illustrated in Fig.\ \ref{fig_HRRIS_architecture}(b), the number and positions of the active elements can be cast as design parameters. In particular, the problem of SE maximization for the dynamic HR-RIS can be formulated as
	\begin{subequations}
		\begin{align*}
			(\mathrm{P1}) \quad \underset{\{\alpha_n\}, \setA}{\textrm{maximize}} \quad &  f_0(\{\alpha_n\}) \nbthis \\
			\textrm{subject to} \quad &\eqref{opt_hyb_con_unit_modul}-\eqref{opt_hyb_con_relay_power}, \nbthis\\
			& \abs{\setA} \leq K, \setA \subset \{1,2,\ldots,N\}, \nbthis \label{con_abs_A}
		\end{align*}
	\end{subequations}
	where the constraint \eqref{con_abs_A} means that the number of active elements in the HR-RIS does not exceed the maximum number of RF-PA chains.  Problem $(\mathrm{P1})$ does not only inherit the numerical challenge of $(\mathrm{P0})$ but also includes the difficulty from cardinality constraint \eqref{con_abs_A}. We develop an efficient solution to $(\mathrm{P1})$ for practical use below.
	%	\subsubsection{Efficient Solution to $(\mathrm{P1})$}
	
	Similar to problem $(\mathrm{P0})$, firstly, we employ the upper bound of $ f_0(\{\alpha_n\})$ (see \eqref{eq_ub_4}) to obtain an approximate but more tractable problem of $(\mathrm{P1})$, which is given as 
	\begin{subequations}
		\begin{align*}
			\pdyn \quad \underset{\{\alpha_n\}, \setA}{\textrm{maximize}} \quad &  f(\{\alpha_n\}) \nbthis \\
			\textrm{subject to} \quad &\eqref{opt_hyb_con_unit_modul}-\eqref{opt_hyb_con_relay_power}, \nbthis\\
			& \abs{\setA} \leq K, \setA \subset \{1,2,\ldots,N\}\nbthis  \label{con_cardinality}
		\end{align*}
	\end{subequations}
	
	At this point, a naive approach to solve $\pdyn$ would be based on an exhaustive search, i.e., using the result presented in Section \ref{sec_efficient_solution} to determine the coefficients $\{\alpha_n\}$ corresponding to each valid set $\setA$, then choosing the one providing the best performance. Clearly, such an approach requires high computational cost due to excessively large numbers of combinations of $\setA$. 
	
	We now present an efficient solution to $\pdyn$. Recall that we can write $\an = \abs{\an} e^{j \theta_n}$. Based on this, let us introduce two matrices as $\bTheta = \diag{e^{j \{ \theta_1 \}}, \ldots, e^{j  \{ \theta_N \}}}$  and $\hat{\bUpsilon} = \diag{\abs{\alpha_1}, \ldots, \abs{\alpha_N}}$ with $\abs{\an} = 1, n \notin \setA$. Then we can write  $\bUpsilon =\bTheta \hat{\bUpsilon} $. This motivates us develop an alternating procedure applied on three sets of variables $\{\bTheta, \hat{\bUpsilon}, \setA\}$. Next, we describe the approach determining each of $\{\bTheta, \hat{\bUpsilon}, \setA\}$ when the others are fixed, which are the key steps in our proposed solution.
	
	\subsubsection{Determine $\{\theta_n\}$}
	\label{sec_dyn_step1}
	For this step, we suppose that $\abs{\an} = 1, \forall n$, which corresponds to the conventional RIS where all elements are passive. Then, we determine $\{\theta_n\}$ via solving the following problem extracted from $\pdyn$
	\begin{align*}
		\pdyna \quad \underset{\{\alpha_n\}}{\textrm{maximize}} \quad &  f(\{\alpha_n\}) \\
		\textrm{subject to} \quad & \abs{\an}=1, \forall n.
	\end{align*}
	A solution to $\pdyna$ can be obtained by using the same approach presented in Section \ref{sec_sol_fixHR} (i.e., see \eqref{eq_alpha_best}) for the case $n \notin \setA$. 
	
	\subsubsection{ Determine $\setA$ and $\{\abs{\an}\}$}
	\label{sec_dyn_step2}
	We recall that the passive reflecting elements have unit modulus, i.e., $\abs{\an}  = 1, \forall n \notin \setA$. Thus, we only need to determine $\abs{\an}, \forall n \in \setA$. Furthermore, an active element only results in performance gain with respect to a passive element if $\abs{\an} > 1$; otherwise, it causes performance loss due to signal attenuation. Therefore, to minimize the loss, in the dynamic HR-RIS architecture, the RF-PA chain connected to the $n$th element should be turned off if $\abs{\an} < 1$. As a result, in the dynamic HR-RIS scheme, we have $\abs{\an} > 1, \forall n \in \setA$, and the solution in \eqref{eq_alpha_best} becomes
		\begin{align*}
			\ans =
			\begin{cases}
				\abs{\ans} e^{-j \arg \{ \lambda_n \}}, & \abs{\ans} > 1, n \in \setA\\
				e^{-j \arg \{ \lambda_n \}}, & \text{otherwise}
			\end{cases},
		\end{align*}
		with $\{\arg \{ \lambda_n \}\}$ being determined in the previous step. With this solution, it is obvious that $\re{\ans \lambda_n} = \abs{\ans} \abs{\lambda_n}$, and \eqref{eq_fn} can be rewritten as
		\begin{align*}
			g_n(\ans) &= \log_2 \left[ \left(1 + \abs{\ans}^2  \gamma_n \right) \right. \\
			&\left. \quad \quad \times  \left(1 + \abs{\ans}^2 \abs{\lambda_n}^2  + 2 \abs{\ans} \abs{\lambda_n} - v_n^{\prime} v_n \abs{\lambda_n}^2 \right) \right]\\
			&= \log_2 \left[ 1 +  \zeta_n^2 \abs{\ans}^4 + o\left(\abs{\ans}^3\right) \right], \nbthis \label{eq_fn1}
		\end{align*} 
		where $\zeta_n \triangleq \abs{\lambda_n} \sqrt{\gamma_n}$, and $ o\left(\abs{\ans}^3\right)$ represents the sum of the remaining terms associated with $\abs{\ans}^c, c \leq 3$. Because $\abs{\an} > 1, n \in \setA$, we can ignore $o\left(\abs{\ans}^3\right)$ in \eqref{eq_fn1} and approximately maximize $g_n(\alpha_n)$ via maximizing $\hat{g}_n(\ans) \triangleq \log_2 \left(1 +  \zeta_n^2 \abs{\ans}^4 \right)$. Moreover, because $\hat{g}_n(\ans)$ monotonically increases with $\abs{\ans}$, maximizing it is equivalent to maximizing $\hat{\hat{g}}_n(\ans) \triangleq \log_2 \left(1 +  \zeta_n \abs{\ans}^2 \right)$. The observation implies that to find $\abs{\ans}$ that maximizes $g_n(\alpha_n^{\star}), \forall n \in \setA$, we can consider the following problem
		\begin{subequations}
			\label{pdyn}
			\begin{align*}
				\underset{\setA,\{\abs{\an}^2\}}{\textrm{maximize}}\quad & \sum_{n \in \setA} \log_2 \left(1 + \zeta_n \abs{\an}^2 \right) \nbthis \label{opt_f_obj_dyn} \\
				\textrm{subject to} \quad & \sum_{n \in \setA} \abs{\an}^2 \xi_n \leq \Pamax \text{ and } \eqref{con_cardinality}, \nbthis \label{opt_f_con_active_dyn}
			\end{align*}
		\end{subequations}
		where the first constraint in \eqref{opt_f_con_active_dyn} is the transmit power constraint of the active elements. Let $p_n \triangleq \abs{\an}^2 \xi_n$ denote the power allocated to the $n$th active element, $n \in \setA$. The problem in \eqref{pdyn} can be rewritten as
		\begin{subequations}
			\begin{align*}
				\underset{\setA,\{p_n\}}{\textrm{maximize}} \quad & \sum_{n \in \setA} \log_2 \left(1 + \frac{\zeta_n}{\xi_n} p_n   \right) \nbthis \label{opt_f_obj_dyn_1} \\
				\textrm{subject to} \quad & \sum_{n \in \setA} p_n \leq \Pamax  \text{ and } \eqref{con_cardinality}, \nbthis \label{opt_f_con_active_dyn_1}
			\end{align*}
		\end{subequations}
		which allows $\setA$ and $\{p_n\}$ to be obtained. Specifically, it is clear from \eqref{opt_f_obj_dyn_1} that the optimal set of active elements, i.e., $\setA^{\star}$, is the set of indices of the $K$ largest elements in $\left\{ \frac{\zeta_1 }{\xi_1}, \ldots, \frac{\zeta_N }{\xi_N} \right\}$, where $\frac{\zeta_n }{\xi_n}, \forall n$ are obtained during determining $\{\theta_n\}$ in the previous step, i.e., in Section \ref{sec_dyn_step1}.
	
	Once $\setA^{\star}$ is determined, $\{p_n\}$ can be solved by performing power allocation to the active elements in $\setA^{\star}$. The optimal solution is the well-know water-filling solution $p_n^{\star} = \max \left\{ \frac{1}{\mu} - \frac{\xi_n}{\zeta_n}, 0 \right\}$, with $\mu$ satisfying \eqref{opt_f_con_active_dyn_1}. As a result, we obtain $\abs{\ans}^2 = \frac{p_n^{\star}}{\xi_n} = \max \left\{ \frac{1}{\mu \xi_n} - \frac{1}{\zeta_n}, 0 \right\}$, $n \in \setA^{\star}$. Consequently, the amplitudes of elements in the dynamic HR-RIS scheme is given as
	\begin{align*}
		\abs{\ans} = 
		\begin{cases}
			\max \left\{ \sqrt{\frac{1}{\mu \xi_n} - \frac{1}{\zeta_n}}, 1 \right\}, &n \in \setA^{\star} \\
			1, & \text{otherwise}
		\end{cases}
		. \nbthis \label{eq_opt_ampl_dyn}
	\end{align*}
	
	The proposed procedure finding efficient solution to the problem of SE maximization in dynamic HR-RIS architecture is outlined in Algorithm \ref{alg_dynamic}. It starts with randomly generating $\{\alpha_n\}$ with unit modules. Then steps 2--9 are executed to determine $\{\theta_n^{\star}\}$. After that, $\setA^{\star}$ is determined in steps 10 and 11, allowing $\{\abs{\ans}\}$ to be obtained in step 12. Finally, the coefficients of the dynamic HR-RIS are derived in step 13 based on \eqref{eq_alpha_best}.
	
	\begin{algorithm}[t]
		%\small
		\caption{Find coefficients of the dynamic HR-RIS}
		\label{alg_dynamic}
		\begin{algorithmic}[1]
			\REQUIRE $\mHt, \mHr$.
			\ENSURE $\setA^{\star},\{{\ans}\}$.
			\STATE Randomly generate $\bTheta = \diag{e^{j \theta_1}, \ldots, e^{j \theta_N}}$ and assign $\bUpsilon = \bTheta$.
			\WHILE{objective value does not converge}
			\FOR {$n = 1 \rightarrow N$}
			\STATE Compute $\mA_n, \mB_n$, and $\mC_n$ based on \eqref{def_A_passive}-\eqref{def_C_passive}.
			\STATE $\mD_n = \mIr + \mA_n^{-1}\mB_n$, $\mE_n = \mA_n \mD_n$.
			\STATE Find $\lambda_n$ and $\gamma_n$ as the sole non-zero eigenvalue of $\mE_n^{-1}\mC_n$ and  $\mA_n^{-1}\mB_n$, respectively.
			\STATE $\theta_n^{\star} = -\arg \{ \lambda_n \}$, $\ans = e^{-j \theta_n^{\star}}$.
			
			\ENDFOR 
			\ENDWHILE
			\STATE Compute $\frac{\zeta_n}{\xi_n}$, $n=1,\ldots,N$, where $\zeta_n \triangleq \abs{\lambda_n} \sqrt{\gamma_n}$ and $\xi_n \triangleq \sigma^2 + \Pbs\norm{\vt_n}^2$.
			\STATE Set $\setA^{\star}$ to the indices of the $K$ largest value in $\left\{ \frac{\zeta_1 }{\xi_1}, \ldots, \frac{\zeta_N }{\xi_N} \right\}$. 
			
			%\STATE Perform the water-filling algorithm to obtain $p_n^{\star}, n \in \setA$.
			\STATE Obtain $\{\abs{\ans}\}$ given in \eqref{eq_opt_ampl_dyn}.
			\STATE Obtain $\{\ans\}$ based on \eqref{eq_alpha_best}.
			%\STATE $\ans = \abs{\ans} e^{-j \arg \{ \lambda_n \}}, \forall n \in \setA^{\star}$. 
		\end{algorithmic}
	\end{algorithm}
	
    \vspace{-0.2cm}
	\section{Power Consumption Analysis}
	\label{sec_power}
	
	\subsection{Fixed HR-RIS Architecture}
	In the fixed HR-RIS, the number of active and passive elements are fixed to $K$ and $M$, respectively. Therefore, the total power consumption of the whole MIMO communications system aided by the fixed HR-RIS can be modeled as \cite{zappone2013energy, gong2019robust, bjornson_intelligent_2019}
	\begin{align}
		\label{eq_p_HRRIS_fixed}
		P_{{\mathrm{H}}}^{\mathrm{fix.}} = \frac{\Pbs}{\tau_{\mathrm{BS}}} + \frac{\Pa}{\tau_{\mathrm{a}}} + P_{\mathrm{c,H}}^{\mathrm{fix.}} + M P_{\mathrm{p}},
	\end{align}
	where $\tau_{\mathrm{BS}}, \tau_{\mathrm{a}} \in (0,1]$ are the power amplifier efficiencies of the BS and active elements at the HR-RIS, respectively, $\Pa$ is given in \eqref{eq_relay_power}, and $P_{\mathrm{p}}$ is the power required for a passive reflecting element \cite{bjornson_intelligent_2019}. Furthermore, $P_{\mathrm{c,H}}^{\mathrm{fix.}}$ is the total circuit power consumption of the fixed HR-RIS-aided MIMO system and can be computed as \cite{zappone2013energy}
	\begin{align*}
		P_{\mathrm{c,H}}^{\mathrm{fix.}} = N_t P_{\mathrm{BS,dynamic}} + K P_{\mathrm{a,dynamic}} + P_{\mathrm{BS,static}} + P_{\mathrm{a,static}} ,% \nbthis \label{eq_p_circuit_HRRIS_fixed}
	\end{align*}
	with $P_{\mathrm{BS,dynamic}}$ denoting the dynamic power consumption of each RF chain of the BS, and $P_{\mathrm{BS,static}}$ denoting the static power overhead of the BS, including baseband processing, power suply, and cooling power consumption. Similarly, $P_{\mathrm{a,dynamic}}$ and $P_{\mathrm{a,static}}$ are the dynamic and static power consumption of the active relay elements in the HR-RIS.
	
	\subsection{Dynamic HR-RIS Architecture}
	In the dynamic HR-RIS architecture, the number of active and passive elements vary depending on $\Pamax$ and $\xi_n$ according to \eqref{eq_opt_ampl_dyn}. Let $\abs{\setA^{\star}}$ denote the number of actual active elements whose amplitudes are larger than one, where $\setA^{\star}$ is obtained in Algorithm \ref{alg_dynamic}. As a result, the number of passive elements is given as $N - \abs{\setA^{\star}}$, and the total power consumption of a dynamic HR-RIS-assisted MIMO system is given by
	\begin{align}
		\label{eq_p_HRRIS_dyn}
		P_{{\mathrm{H}}}^{\mathrm{dyn.}} = \frac{\Pbs}{\tau_{\mathrm{BS}}} + \frac{\Pa}{\tau_{\mathrm{a}}} + P_{\mathrm{c,H}}^{\mathrm{dyn.}} + (N - \abs{\setA^{\star}}) P_{\mathrm{p}} + N P_{\mathrm{SW}},
	\end{align}
	where the last term accounts for the total power consumed by the $N$ switches in Fig.\ \ref{fig_HRRIS_architecture}(b), each requires a power of $P_{\mathrm{SW}}$, and $P_{\mathrm{c,H}}^{\mathrm{dyn.}}$ is the total circuit power consumption of the dynamic HR-RIS-aided MIMO system, given as \cite{zappone2013energy}
	\begin{align*}
		P_{\mathrm{c,H}}^{\mathrm{dyn.}} &= N_t P_{\mathrm{BS,dynamic}} + \abs{\setA^{\star}} P_{\mathrm{a,dynamic}} \\
		&\hspace{3cm}+ P_{\mathrm{BS,static}} + P_{\mathrm{a,static}}.
	\end{align*}

	\subsection{Conventional RIS Architecture}
	\label{sec_power_compare}
	We compare the total power consumption of the proposed HR-RIS-assisted system to that of the conventional RIS-assisted system. We note that in the latter, there is no active relay element, but $N = M + K$ passive reflecting elements are used at the RIS. Therefore, the total power consumption of the RIS-aided system can be expressed as
	\begin{align}
		\label{eq_p_RIS}
		P_{\mathrm{RIS}} = \frac{\Pbs}{\tau_{\mathrm{BS}}} + P_{\mathrm{c,RIS}} + NP_{\mathrm{p}},
	\end{align}
	where $P_{\mathrm{c,RIS}} = N_t P_{\mathrm{BS,dynamic}} + P_{\mathrm{BS,static}}$ is the total circuit power consumption of the dynamic HR-RIS-aided MIMO system
	
	By comparing \eqref{eq_p_HRRIS_fixed} and \eqref{eq_p_HRRIS_dyn} to \eqref{eq_p_RIS}, with the note that $P_{\mathrm{passive}} \ll P_{\mathrm{a,dynamic}}$, the increased total power consumption of the fixed and dynamic HR-RIS-aided systems with respect to the RIS-assisted system can be respectively given as 
	\begin{align*}
		\Delta P^{\mathrm{fix.}}_{\mathrm{H}} = \frac{\Pa}{\tau_{\mathrm{a}}} + K \left(P_{\mathrm{a,dynamic}} - P_{\mathrm{p}}\right) + P_{\mathrm{a,static}},\\
		\Delta P^{\mathrm{dyn.}}_{\mathrm{H}} = \frac{\Pa}{\tau_{\mathrm{a}}} + \abs{\setA^{\star}} \left(P_{\mathrm{a,dynamic}} - P_{\mathrm{p}}\right) + P_{\mathrm{a,static}},% \nbthis \label{eq_deltaP_dyn}
	\end{align*}
	which linearly increases with the number of active elements, i.e., $K$ in the fixed HR-RIS and $\abs{\setA^{\star}}$ in the dynamic HR-RIS. The numerical comparison of these architectures will be presented in the next section.
	
	%========================= Simulation results ==============================================
	\section{Simulation Results}
	\label{sec_sim}
	
	\begin{figure}[t]
		\centering
		\belowcaptionskip = -0.5cm
		\includegraphics[scale=0.5]{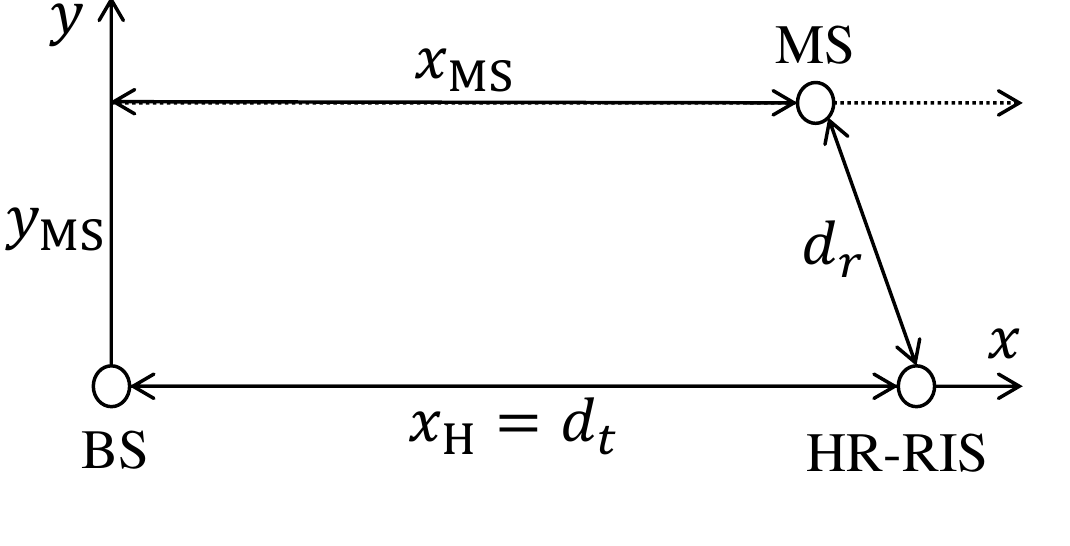}
		\caption{Horizontal locations of the BS, MS, and HR-RIS.}
		\label{fig_system_illustration}
	\end{figure}
	
	In this section, numerical results are provided to validate the proposed HR-RIS schemes. We assume that uniform linear arrays (ULAs) are deployed at the BS and MS, respectively. By contrast, an uniform planar array (UPA) of $N$ elements is employed for the HR-RIS. Furthermore, half-wavelength distancing between array elements are assumed for the BS, MS, and HR-RIS. As shown in Fig.\ \ref{fig_system_illustration}, in a two-dimensional coordinate, we assume that the BS is deployed at a fixed location $(0,0)$. By contrast, the HR-RIS and MS are located at $(x_{\mathrm{H}},0)$ and $(x_{\mathrm{MS}},y_{\mathrm{MS}})$, respectively, where $x_{\mathrm{H}}$ and $x_{\mathrm{MS}}$ can vary. As a result, the distance between the BS and MS is given by $d_t = x_{\mathrm{H}}$, while that between the HR-RIS and the MS is $d_r = \sqrt{(x_{\mathrm{H}} - x_{\mathrm{MS}})^2 + y_{\mathrm{MS}}^2}$. The path loss of a link distance $d$ is given by \cite{wu2019intelligent, zhang2020capacity} $\beta(d) = \beta_0 \left(\frac{d}{1\mathrm{ m}}\right)^{\epsilon}$, where $\beta_0$ is the path loss at the reference distance of $1$ meter (m), and $\epsilon$ is the path loss exponent.
	
	For small-scale fading, we assume the Rician fading channel model \cite{wu2019intelligent, zhang2020capacity}. As a result, the small-scale fading channel between the BS and the HR-RIS is modeled by \cite{wu2019intelligent, zhang2020capacity}
	\begin{align*}
		\tilde{\mH}_t = \left(\sqrt{\frac{\kappa_t}{1+\kappa_t}} \mHt^{\mathrm{LoS}} + \sqrt{\frac{1}{1+\kappa_t}} \mHt^{\mathrm{NLoS}}\right), \nbthis \label{eq_Ht}
	\end{align*}
	where $\mHt^{\mathrm{LoS}}$ and $\mHt^{\mathrm{NLoS}}$ represent the deterministic LoS and non-LoS (NLoS) components, respectively. The NLoS channel is modeled by the Rayleigh fading, with the entry on the $i$th row and $j$th column of $\mHt^{\mathrm{NLoS}}$ being given as $h_{t,ij}^{\mathrm{NLoS}} \sim \mathcal{CN} (0,1)$. The LoS component for each channel is modeled as a deterministic component, i.e., $\mHt^{\mathrm{LoS}} =  \va_{\mathrm{H}} (\theta_{\mathrm{H}},\phi_{\mathrm{H}}) \va_{\mathrm{BS}}^H (\theta_{\mathrm{BS}})$, where $\va_{\mathrm{BS}} (\theta_{\mathrm{BS}})$ and $\va_{\mathrm{H}} (\theta_{\mathrm{H}},\phi_{\mathrm{H}})$ are the array response vectors at the BS and HR-RIS, respectively. Here, the $n$th element of $\va_{\mathrm{BS}} (\theta_{\mathrm{BS}})$ and $\va_{\mathrm{H}} (\theta_{\mathrm{H}},\phi_{\mathrm{H}})$ are given as  $a_{\mathrm{BS},n} (\theta_{\mathrm{BS}}) = e^{j \pi (n-1) \sin \theta_{\mathrm{BS}}}, n = 1, \ldots, N_t$ and $a_{\mathrm{H},n} (\theta_{\mathrm{H}},\phi_{\mathrm{H}}) = e^{j \pi \left( \lfloor \frac{n}{N_x} \rfloor \sin \theta_{\mathrm{H}} \sin \phi_{\mathrm{H}} + \left( n - \lfloor \frac{n}{N_x} \rfloor N_x \right) \sin \theta_{\mathrm{H}} \cos \phi_{\mathrm{H}} \right) }$, $n = 1, \ldots, N$ \cite{zhang2020capacity}. $\theta_{\mathrm{BS}}$, $\theta_{\mathrm{H}} \in [0, 2\pi)$ denote the angle-of-departure (AoD) at the BS and the azimuth angle-of-arrival (AoA) at the HR-RIS, respectively, and $\phi_{\mathrm{H}} \in [-\pi / 2, \pi / 2)$ denotes the elevation AoA at the HR-RIS. Furthermore, in \eqref{eq_Ht}, $\kappa_t$ is the Rician factor. With $\kappa_t \rightarrow 0$, $\tilde{\mH}_t$ approaches the Rayleigh fading channel, and with $\kappa_t \rightarrow \infty$, $\tilde{\mH}_t$ becomes the LoS channel. The channel matrix between the BS and the HR-RIS, i.e., $\mHt$, is obtained by multiplying $\tilde{\mH}_t$ by the square root of the corresponding path loss $\beta(d_t)$, i.e., $\mHt = \sqrt{\beta(d_t)} \tilde{\mH}_t$. The channel matrix between the HR-RIS and MS is modeled similarly.
	
	In this work, unless otherwise stated, we set $\beta_0 = -30$ dB, $\{\epsilon_t,\epsilon_r\} = \{2.2,2.8\}$,  $\sigma^2 = -80$ dBm, $\{x_{\mathrm{MS}}, y_{\mathrm{MS}}, x_{\mathrm{H}} \} = \{40,2,51\}$ m, and $\{\kappa_t,\kappa_r\} = \{\infty,0\}$ \cite{wu2019intelligent}. Furthermore, the resolution of the phase shifts is set to $b= 2$ bits, i.e., the phase shifts are selected from the set $\mathcal{F} = \left\{\frac{2\pi (q-1)}{Q}\right\}$, $q=1,\ldots, Q$, where $Q = 2^b$. All the numerical results are averaged over $100$ independent channel realizations.
	
	\subsection{Improvement in SE of the Proposed HR-RIS}
	
	\begin{figure}[t]
		\centering
		\vspace{-0.5cm}
		\belowcaptionskip = -10pt
		\includegraphics[scale=0.55]{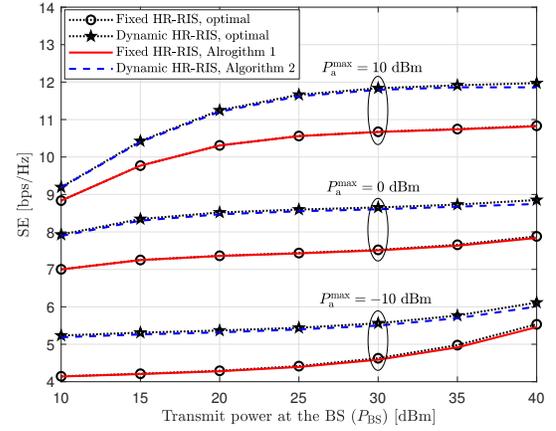}
		\caption{SEs of the proposed HR-RIS schemes compared to those obtained by the exhaustive search for $N_t = 4$, $N_r = 2$, $N = 4$, $K = 1$, $M = N-K=3$, and $\Pamax = \{-10, 0, 10\}$ dBm.}
		\label{fig_rate_vs_Pt_ES}
	\end{figure}
	
	\begin{figure}[t]
		\centering
		\belowcaptionskip = -15pt
		\includegraphics[scale=0.55]{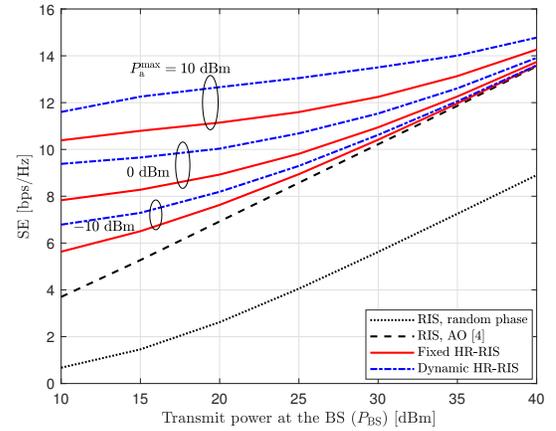}
		\caption{SEs of the proposed HR-RIS schemes compared to those of the conventional RIS schemes for $N_t = 32$, $N_r = 2$, $N = 50$, $K = 1$, $M = N-K=49$, and $\Pamax = \{-10, 0, 10\}$ dBm.}
		\label{fig_rate_vs_Pt}
	\end{figure}
	
	We at first consider in Fig.\ \ref{fig_rate_vs_Pt_ES} a $4 \times 2$ MIMO system, where a small HR-RIS with $N=4$ and $K=1$ is deployed, to show that the proposed schemes can perform near-optimal. With $K=1$, the HR-RIS has only a single active element, whose transmit power is set to $\Pamax=\{-10, 0, 10\}$ dBm. To obtain the optimal performance, we perform the exhaustive search over all the possible solutions. Specifically, for the fixed HR-RIS, the index of the active element is set to one, and the optimal solution is searched over all combinations of $N$ phase values, each is in $\mathcal{F}$. For the dynamic HR-RIS, the exhaustive search scheme searches for the optimal combination of $N$ phase values and the index of the active element. It is observed from Fig.\ \ref{fig_rate_vs_Pt_ES} that the proposed schemes perform very close to the optimal exhaustive search counterparts for all the considered values of $\Pamax$. In the subsequent simulations, we consider larger systems and omit the results for exhaustive search because they require excessively high computational and time complexity.

	In Fig.\ \ref{fig_rate_vs_Pt}, we show the performance improvement of the proposed HR-RIS architectures in terms of SE. Specifically, a $32 \times 2$ MIMO system aided by an intelligent surface of $N=50$ elements is considered. The SEs of the proposed fixed and dynamic HR-RIS schemes are shown for $K=1$, $M=N-K=49$, and $\Pamax=\{-10, 0, 10\}$ dBm. For comparison, we also show the SEs of the conventional RIS with $N$ fully passive reflecting elements, whose phases are either randomly generated or optimized using the AO method in \cite{zhang2020capacity}. Furthermore, in all the simulation results for the fixed HR-RIS scheme, $\setA$ is fixed to $\{1,\ldots,K\}$. It is noted that the coefficients of the HR-RIS and AO-based RIS \cite{zhang2020capacity} are obtained based on the same optimization method, i.e., the AO approach.

	It can be observed from Fig.\ \ref{fig_rate_vs_Pt} that the RIS with random phases has poor performance compared to the optimized RIS. By contrast, the HR-RIS schemes achieve significant improvement in the SE with respect to the conventional RIS, especially at low $\Pbs$. This numerically justifies the observation in Remark \ref{rm_gain_Pt}, which states that a higher performance gain can be obtained by the HR-RIS for low $\Pbs$. Furthermore, it is seen that the HR-RIS only requires a small $\Pamax$ to achieve remarkable SE improvement at low $\Pbs$, and with $\Pamax = \{0, 10\}$ dBm, sustainable SE gains of the HR-RIS are seen for the whole considered range of $\Pbs$. For example, the fixed and dynamic HR-RIS only requires $\Pamax = 0$ dBm to save more than $10$ dBm and $15$ dBm for the transmit power at the BS, respectively. However, with a small power budget at the HR-RIS, but a large transmit power at the BS, e.g., with $\Pamax = -10$ dBm and $\Pbs \geq 35$ dBm, the performance gain of the HR-RIS decreases, and it performs comparably with the conventional RIS. This observation agrees with the discussion in Remarks \ref{rm_gain_vs_K} and \ref{rm_gain_Pt}. The SE gains of the proposed schemes for low $\Pbs$ have been clearly shown in this figure. Therefore, in the following results on SE, we consider high values of $\Pbs$.

	\begin{figure}[t]
		\centering
		\vspace{-0.4cm}
		\belowcaptionskip = -15pt
		\includegraphics[scale=0.55]{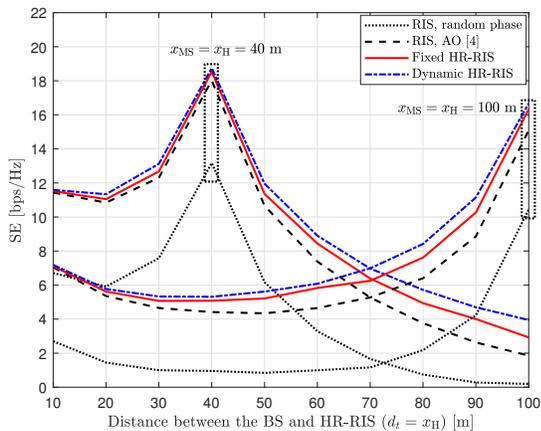}
		\caption{SE improvement for different positions of the HR-RIS with $N_t = 32$, $N_r = 2$, $K = 1$,  $\Pbs = 30$ dBm, and $\Pamax = 0$ dBm.}
		\label{fig_rate_vs_d}
	\end{figure}
	
	In Fig.\ \ref{fig_rate_vs_d}, we investigate the SEs of the HR-RIS for different geographical deployments of the RIS/HR-RIS by varying $x_{\mathrm{H}}$. We note that for comparison, the RIS and HR-RIS are set to be placed at the same position. In this simulation, we set $N_t = 32$, $N_r = 2$, $K = 4$,  $\Pbs = 30$ dBm, and $\Pamax = 0$ dBm. The BS is placed at the coordinate $(0,0)$, while two positions are considered for the MS, including $\{(40,2), (100,2)\}$. The HR-RIS/RIS is placed at $(0, x_{\mathrm{H}})$, where $x_{\mathrm{H}} \in [10, 100]$ m, implying that the HR-RIS/RIS moves along the $x-$axis in Fig.\ \ref{fig_system_illustration}. An observation from Fig.\ \ref{fig_rate_vs_d}, which agrees with the finding in \cite{bjornson_intelligent_2019}, is that both the RIS and HR-RIS have decreased SEs when they move far away from either the BS or the MS, and the highest SEs are attained when they are closest to the MS. However, in this figure, we focus more on the SE gains of the proposed HR-RIS versus $x_{\mathrm{H}}$. Specifically, it is seen that as $x_{\mathrm{H}}$ grows, path loss $\beta (d_t)$ increases, leading to a smaller $\xi_n$ in \eqref{eq_optimal_active_amplitude} and \eqref{eq_opt_ampl_dyn}. As result, more significant active beamforming gains are attained, as discussed in Remark \ref{rm_gain_Pt}. Furthermore, the dynamic HR-RIS exhibits large gains over the others in the worst scenarios, i.e., when the MS is far from both the HR-RIS/RIS and BS. It can be concluded from Fig.\ \ref{fig_rate_vs_d} that the proposed HR-RIS is robust over geographical deployment.
	
	\begin{figure*}[t]
		%\vspace{-0.5cm}
		\belowcaptionskip = -15pt
		\centering
		\includegraphics[scale=0.6]{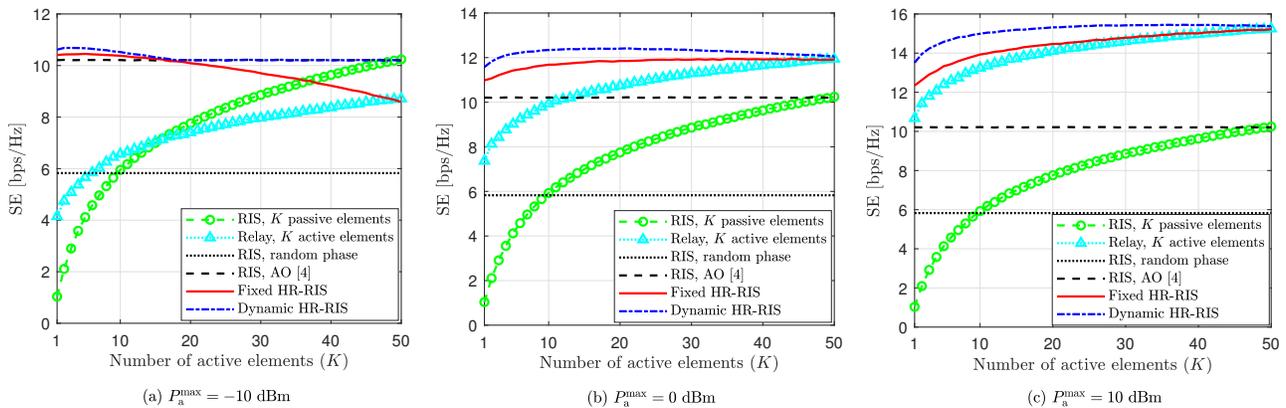}
		\caption{SEs of the proposed HR-RIS schemes compared to those of the conventional RIS schemes and the relay with $K$ active elements for $N_t = 32$, $N_r = 2$, $N = 50$, $K = \{1,2,\ldots,N\}$, $\Pbs = 30$ dBm, and $\Pamax = \{-10, 0, 10\}$ dBm.}
		\label{fig_rate_vs_K}
	\end{figure*}
	
	Furthermore, we investigate the SE improvement of the proposed schemes for different values of $K$ in Fig.\ \ref{fig_rate_vs_K}. In this figure, the proposed schemes are compared to not only the conventional RIS with random phases and AO-based RIS, which are equipped with $N$ elements, but also to the RIS and the fully-active relay equipped with only $K$ independent elements. We consider $\Pbs = 30$ dBm, $\Pamax = \{-10, 0, 10\}$ dBm, and the other simulation parameters are the same as those in Fig.\ \ref{fig_rate_vs_Pt}. The following observations are noted:
	\begin{itemize}
		\item Clearly, the SEs of the conventional fully-passive RIS schemes are constant with $K$. Considering the RIS and relay with only $K$ elements, their SEs monotonically increase with $K$. Furthermore, it is seen that as $K$ increases, the RIS performs closer to the relay with the same number of elements. In particular, a sufficiently large RIS can outperform the relay with a limited power budget, as seen in Fig.\ \ref{fig_rate_vs_K}(a). These observations agree with the findings in \cite{bjornson_intelligent_2019} and \cite{wu2019intelligent}. However, for a small $K$, the relay performs far better than the RIS, even with a limited power budget, which has motivated the proposal of the HR-RIS in this work.
		
		\item Comparing the SEs of the proposed HR-RIS and the conventional RIS with $N$ elements, in all the considered scenarios, the former only requires a single active element to outperform the latter. However, as we discussed in Remark \ref{rm_gain_vs_K}, it is clear for the fixed HR-RIS that increasing $K$ does not guarantee the SE improvement, especially for low $\Pamax$ in Fig.\ \ref{fig_rate_vs_K}(a). More specifically, as $K$ increases, the SEs of both the fixed and dynamic HR-RIS quickly increases for small $K$, then they gradually reach their peaks at a small or moderate $K$ before degradation. Therefore, we can conclude that a small number of active elements is sufficient for the HR-RIS to achieve significant improvement in SE with respect to the conventional RIS.
		
		\item The advantage of the dynamic HR-RIS in terms of SE compared to the fixed HR-RIS can be clearly seen in Fig.\ \ref{fig_rate_vs_K}. Specifically, unlike the fixed HR-RIS, in all the considered scenarios, dynamic HR-RIS achieves higher or the same SE compared to the AO-based RIS. A higher $\Pamax$ provides it with a higher gain, but small $\Pamax$ does not cause any performance loss because the active elements can be deactivated to serve as the conventional reflecting elements if they cannot amplify the signal. In general, it can be concluded that the dynamic HR-RIS is more robust than the fixed one.
	\end{itemize}
	
	\begin{figure}[t]
		\centering
		\belowcaptionskip = -0.25cm
		\includegraphics[scale=0.53]{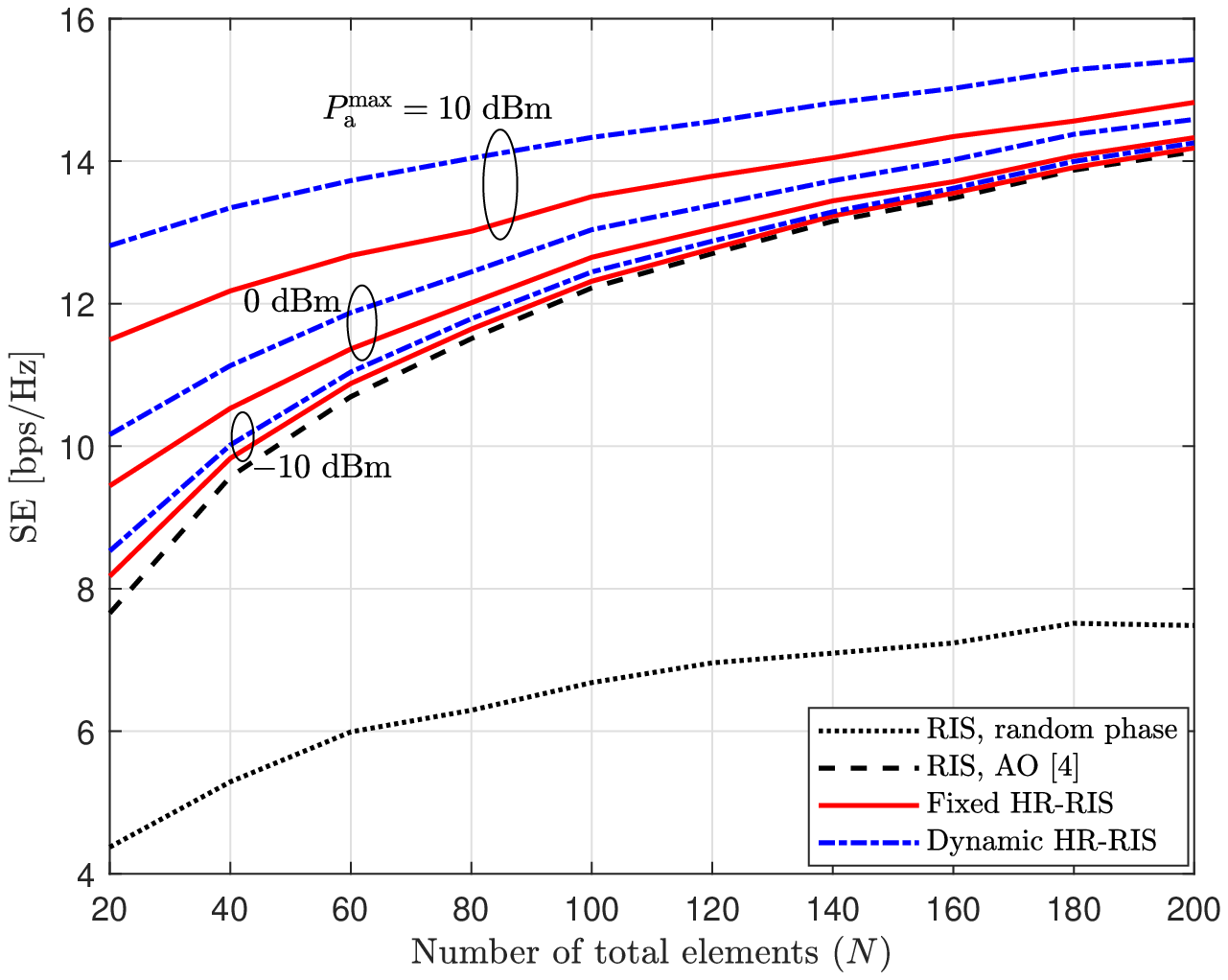}
		\caption{SEs of the proposed HR-RIS schemes compared to those of the conventional RIS schemes for $N_t = 32$, $N_r = 2$, $N \in [20,200]$, $K = 1$, $\Pbs = 30$ dBm, and $\Pamax = \{-10, 0, 10\}$ dBm}
		\label{fig_rate_vs_N}
	\end{figure}

	In Fig.\ \ref{fig_rate_vs_N}, the SEs of the proposed HR-RIS and the conventional RIS schemes are shown for $N_t = 32$, $N_r = 2$, $K = 1$, $\Pbs = 30$ dBm, $\Pamax = \{-10, 0, 10\}$ dBm, and $N \in [20, 200]$. It is seen that the HR-RIS schemes attain the improvement in SE for all the considered values of $N$ with respect to the conventional RIS schemes. In particular, it is observed that the performance improvement becomes less significant as $N$ increases. Nevertheless, the gain of the proposed schemes is still considerable at large $N$, especially with $\Pamax = 10$ dBm. With a larger $\Pamax$, the dynamic HR-RIS achieves higher SE gains with respect to the fixed one. 
	
	\subsection{Power Consumption and EE of the HR-RIS Schemes}
	
	In this subsection, the power consumption and EE of the proposed HR-RIS architectures are investigated and compared to those of the conventional RIS. Specifically, the total power consumption of the MIMO system aided by the HR-RIS schemes and RIS, i.e., $P_{{\mathrm{H}}}^{\mathrm{fix.}}$, $P_{{\mathrm{H}}}^{\mathrm{dyn.}}$, and $P_{\mathrm{RIS}}$, are computed based on \eqref{eq_p_HRRIS_fixed}, \eqref{eq_p_HRRIS_dyn} and \eqref{eq_p_RIS}, respectively. The component power consumption is assumed as follows: $P_{\mathrm{BS,dynamic}} = 40$ dBm, $P_{\mathrm{a,dynamic}} = 35$ dBm, $P_{\mathrm{BS,static}} = 35$ dBm, $P_{\mathrm{a,static}} = 30$ dBm, $P_{\mathrm{p}} = P_{\mathrm{SW}} =5$ mW, and $\tau_{\mathrm{a}} = \tau_{\mathrm{BS}} = 0.5$ \cite{gong2019robust, bjornson_intelligent_2019, nguyen2019unequally}. Then, the EE of a scheme is given as $\mathrm{EE} = \frac{W  \mathrm{SE}}{P}$ [bps/W], where $W$ denotes the system bandwidth, and $P \in \{ P_{{\mathrm{H}}}^{\mathrm{fix.}}, P_{{\mathrm{H}}}^{\mathrm{dyn.}}, P_{\mathrm{RIS}} \}$ is the total power consumption of the HR-RIS or RIS-assisted MIMO system. We consider the system bandwidth of $10$ MHz \cite{zhang2020capacity}.
	
	\begin{figure}[t]
		% Fig.\ 2
		\centering
		\belowcaptionskip = -0.4cm
		\includegraphics[scale=0.53]{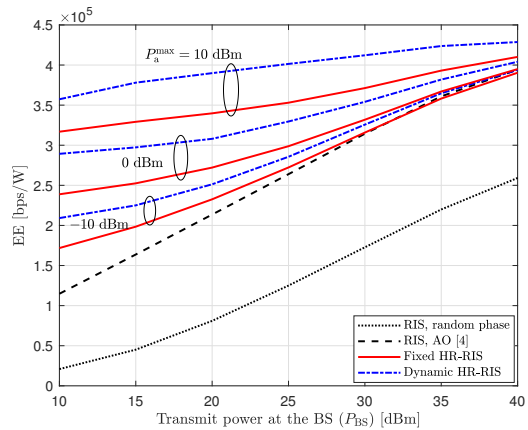}
		\caption{EEs of the proposed HR-RIS schemes compared to those of the conventional RIS schemes for $N_t = 32$, $N_r = 2$, $N = 50$, $K = 1$, $M = N-K=49$, and $\Pamax = \{ -10, 0, 10 \}$ dBm.}
		\label{fig_EE_vs_Pt}
	\end{figure}

	In Fig.\ \ref{fig_EE_vs_Pt}, the EEs corresponding to the SEs in Fig.\ \ref{fig_rate_vs_Pt} are shown. The proposed HR-RIS schemes achieve much higher EEs compared to those of the conventional RIS schemes for low and moderate $\Pbs$. At high $\Pbs$, HR-RIS and RIS have comparable EEs. The EE improvement of the HR-RIS schemes shown in this figure is similar to the SE improvement shown in Fig.\ \ref{fig_rate_vs_Pt}. In the following results on the EE, we focus on comparing the power consumption and EEs of the HR-RIS and RIS for high $\Pbs$ because that for low $\Pbs$ is clear from this figure. In particular, the dynamic HR-RIS will be shown to perform robustly and can save power to ensure high EEs with low $\Pamax$.
	
	\begin{figure*}[t]
		\centering
		%\vspace{-0.5cm}
		\includegraphics[scale=0.6]{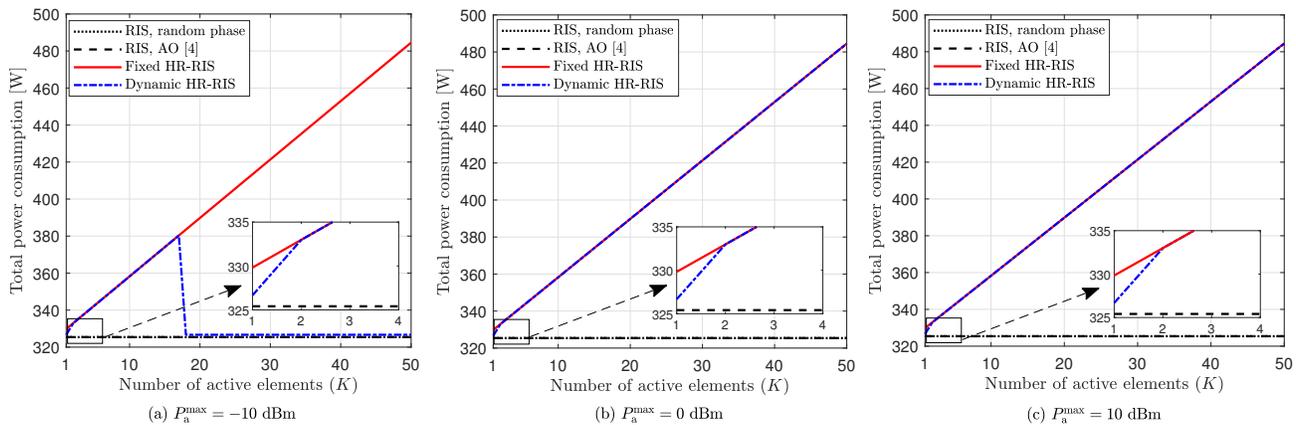}
		\caption{Total power consumption of the proposed HR-RIS schemes compared to those of the conventional RIS schemes for $N_t = 32$, $N_r = 2$, $N = 50$, $K = \{1,2,\ldots,N\}$, $\Pbs = 30$ dBm, and $\Pamax = \{-10, 0, 10\}$ dBm.}
		\label{fig_power_vs_K}
	\end{figure*}

	Assuming the same simulation parameters as those in Fig.\ \ref{fig_rate_vs_K}, we show the power consumption of the considered schemes for $K = \{1, 2, \ldots, N\}$ in Fig.\ \ref{fig_power_vs_K}. As analyzed in Section \ref{sec_power}, it is clear that the proposed HR-RIS architectures require higher power consumption than the RIS. However, for small $\Pamax$, as $K$ increases, the changing properties of $P_{{\mathrm{H}}}^{\mathrm{fix.}}$ and $P_{{\mathrm{H}}}^{\mathrm{dyn.}}$ are of much different. Specifically, for all the considered $\Pamax$,  $P_{{\mathrm{H}}}^{\mathrm{fix.}}$ linearly increases with $K$, which is, however, only correct for the dynamic HR-RIS with $\Pamax = \{ 0, 10 \}$ dBm. In Fig.\ \ref{fig_power_vs_K}(a), $P_{{\mathrm{H}}}^{\mathrm{dyn.}}$ almost linearly increases at first, then quickly drops to a power value that is as low as $P_{\mathrm{RIS}}$. The reason is that, with numerous active elements, $\Pamax = -10$ dBm is not sufficient to share among all the active elements for signal amplifying. Therefore, the dynamic HR-RIS deactivates them to save power and also to preserve the SE, as justified in Fig.\ \ref{fig_rate_vs_K}(a). Focusing on the results for small $K$, e.g., $K = \{1,2,\ldots,4\}$, it is observed from Figs.\ \ref{fig_rate_vs_Pt}, \ref{fig_rate_vs_K}, and \ref{fig_power_vs_K} that the HR-RIS requires a reasonably additional power consumption to attain significant improvement in the SE.
	
	\begin{figure*}[t]
		%\vspace{-15pt}
		\belowcaptionskip = -0.25cm
		\centering
		\includegraphics[scale=0.6]{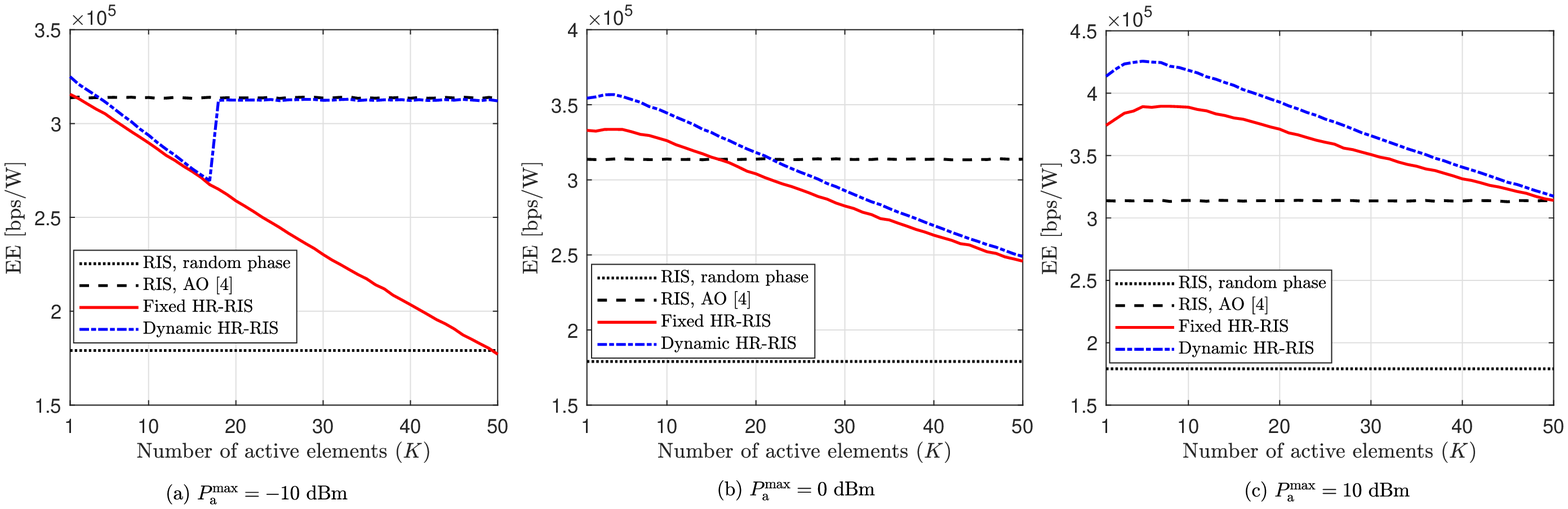}
		\caption{EEs of the proposed HR-RIS schemes compared to those of the conventional RIS schemes for $N_t = 32$, $N_r = 2$, $N = 50$, $K = \{1,2,\ldots,N\}$, $\Pbs = 30$ dBm, and $\Pamax = \{-10, 0, 10\}$ dBm.}
		\label{fig_EE_vs_K}
	\end{figure*}

	Following the investigation of SEs and power consumption of the HR-RIS and RIS in Figs.\ \ref{fig_rate_vs_K} and \ref{fig_power_vs_K}, with the same simulation parameters, we investigate their EEs in Fig.\ \ref{fig_EE_vs_K}. It is seen that the EE of the fixed HR-RIS rapidly decreases as $K$ increases due to its increased power consumption, as shown in Fig.\ \ref{fig_power_vs_K}. However, the EEs of the proposed HR-RIS architectures are the same or much higher than those of the conventional AO-based RIS at small $K$. At large $K$, the dynamic HR-RIS still exhibits the comparable EE with respect to the AO-based RIS for $\Pamax = -10$ dBm. The results in this figure and Fig.\ \ref{fig_rate_vs_K} further show that a spectral- and energy-efficient HR-RIS should be equipped with a small number of active elements. However, if the favorable conditions on $\Pbs$, $\Pamax$, or $K$ are not guaranteed, the dynamic HR-RIS can still manage to ensure no or small loss in EE.

	%========================= Conclusion ====================================================
	\section{Conclusion}
	\label{sec_concusion}
	
	We proposed a novel HR-RIS to assist MIMO communication systems with significantly improved SE and EE compared to that assisted by the conventional RIS. The HR-RIS is a semi-passive beamforming architecture, in which a few elements are capable of adjusting the incident signals' power. It has been designed and optimized by the AO and power allocation strategies, resulting in two different architectures, namely, the fixed and dynamic HR-RIS. In the former, the active elements are fixed; by contrast, those in the latter can be dynamically selected based on their capability of enhancing the SE, and are powered based on the water-filling power allocation method. The analytical results suggest the favorable design and deployment of the HR-RIS. Specifically, it should be equipped with a small number of active elements because employing numerous active elements does not guarantee SE or EE improvement, while consuming high power. Furthermore, the HR-RIS can attain significant SE and EE improvement when the transmit power at the transmitter is small or moderate, and the improvement is more significant when the distance between the transmitter and the HR-RIS is large. Finally, intensive simulations have been performed to numerically justify the findings. The results reveal that the HR-RIS has the potential to outperform both the passive RIS and active AF relaying in certain communications channels.
	
	Several important open problems for future work remain. The practical implementation of the HR-RIS using efficient RAs needs more attention. In particular, the realization of the FD type AF relay assumed has practical challenges. Regardless, this paper provides theoretical insight in bridging the gap between the RIS and active relaying concepts. The system optimization from the EE perspective should be further addressed because we did not directly optimize the EE metric. The CSI acquisition and HR-RIS performance under imperfect CSI, hardware imperfections, and practical wireless channel models need further investigation. Further benefits of HR-RIS can be expected via channel estimation using the active components compared to the passive RIS case. The active components were assumed to be ideal AF relaying with no processing delays similar to the passive reflection. The practical delays and use of DF relaying need to be addressed in future works.
	
	%\appendices

	%========================= References ====================================================
	\bibliographystyle{IEEEtran}
	\bibliography{IEEEabrv,Ref}

\begin{thebibliography}{10}
\providecommand{\url}[1]{#1}
\csname url@samestyle\endcsname
\providecommand{\newblock}{\relax}
\providecommand{\bibinfo}[2]{#2}
\providecommand{\BIBentrySTDinterwordspacing}{\spaceskip=0pt\relax}
\providecommand{\BIBentryALTinterwordstretchfactor}{4}
\providecommand{\BIBentryALTinterwordspacing}{\spaceskip=\fontdimen2\font plus
\BIBentryALTinterwordstretchfactor\fontdimen3\font minus
  \fontdimen4\font\relax}
\providecommand{\BIBforeignlanguage}[2]{{%
\expandafter\ifx\csname l@#1\endcsname\relax
\typeout{** WARNING: IEEEtran.bst: No hyphenation pattern has been}%
\typeout{** loaded for the language `#1'. Using the pattern for}%
\typeout{** the default language instead.}%
\else
\language=\csname l@#1\endcsname
\fi
#2}}
\providecommand{\BIBdecl}{\relax}
\BIBdecl

\bibitem{Alkhateeb2014}
A.~{Alkhateeb}, O.~{El Ayach}, G.~{Leus}, and R.~W. {Heath}, ``Channel
  estimation and hybrid precoding for millimeter wave cellular systems,''
  \emph{{IEEE} J. Sel. Topics Signal Process.}, vol.~8, no.~5, pp. 831--846,
  Oct 2014.

\bibitem{He2014}
J.~{He}, T.~{Kim}, H.~{Ghauch}, K.~{Liu}, and G.~{Wang}, ``Millimeter wave
  {MIMO} channel tracking systems,'' in \emph{Proc. IEEE GLOBECOM Workshop},
  Dec. 2014, pp. 416--421.

\bibitem{Heath2016}
R.~W. {Heath}, N.~{González-Prelcic}, S.~{Rangan}, W.~{Roh}, and A.~M.
  {Sayeed}, ``An overview of signal processing techniques for millimeter wave
  {MIMO} systems,'' \emph{{IEEE} J. Sel. Topics Signal Process.}, vol.~10,
  no.~3, pp. 436--453, April 2016.

\bibitem{zhang2020capacity}
S.~Zhang and R.~Zhang, ``Capacity characterization for intelligent reflecting
  surface aided mimo communication,'' \emph{{IEEE} J. Sel. Areas Commun.},
  vol.~38, no.~8, pp. 1823--1838, 2020.

\bibitem{Wu2019}
Q.~{Wu} and R.~{Zhang}, ``Beamforming optimization for wireless network aided
  by intelligent reflecting surface with discrete phase shifts,'' \emph{{IEEE}
  Trans. Commun.}, pp. 1--1, 2019.

\bibitem{Hu2018_SP}
S.~{Hu}, F.~{Rusek}, and O.~{Edfors}, ``Beyond massive {MIMO}: The potential of
  positioning with large intelligent surfaces,'' \emph{{IEEE} Trans. Signal
  Process.}, vol.~66, no.~7, pp. 1761--1774, Apr. 2018.

\bibitem{Liaskos18}
C.~Liaskos, A.~Tsioliaridou, A.~Pitsillides, S.~Ioannidis, and I.~F. Akyildiz,
  ``Using any surface to realize a new paradigm for wireless communications,''
  \emph{Commun. ACM}, vol.~61, pp. 30--33, 2018.

\bibitem{2019Basar}
E.~{Basar}, M.~{Di Renzo}, J.~{De Rosny}, M.~{Debbah}, M.~{Alouini}, and
  R.~{Zhang}, ``Wireless communications through reconfigurable intelligent
  surfaces,'' \emph{IEEE Access}, vol.~7, pp. 116\,753--116\,773, 2019.

\bibitem{Huang2018}
C.~{Huang}, A.~{Zappone}, G.~C. {Alexandropoulos}, M.~{Debbah}, and C.~{Yuen},
  ``Reconfigurable intelligent surfaces for energy efficiency in wireless
  communication,'' \emph{{IEEE} Trans. Wireless Commun.}, vol.~18, no.~8, pp.
  4157--4170, 2019.

\bibitem{di_renzo_smart_2019}
M.~D. Renzo, M.~Debbah, D.~T.~P. Huy, A.~Zappone, M.~Alouini, C.~Yuen,
  V.~Sciancalepore, G.~C. Alexandropoulos, J.~Hoydis, H.~Gacanin, J.~de~Rosny,
  A.~Bounceur, G.~Lerosey, and M.~Fink, ``Smart radio environments empowered by
  reconfigurable {AI} meta-surfaces: an idea whose time has come,''
  \emph{{EURASIP} J. Wireless Comm. Network.}, vol. 2019, p. 129, 2019.

\bibitem{He2019large}
J.~{He}, H.~{Wymeersch}, L.~{Kong}, O.~{Silv\'en}, and M.~{Juntti}, ``Large
  intelligent surface for positioning in millimeter wave {MIMO} systems,'' in
  \emph{Proc. of IEEE Veh. Tech. Conf. (VTC2020-Spring)}, 2020, pp. 1--5.

\bibitem{taha2019enabling}
A.~Taha, M.~Alrabeiah, and A.~Alkhateeb, ``Enabling large intelligent surfaces
  with compressive sensing and deep learning,'' \emph{arXiv preprint
  arXiv:1904.10136}, 2019.

\bibitem{taha2019deep}
------, ``{Deep learning for large intelligent surfaces in millimeter wave and
  massive MIMO systems},'' in \emph{IEEE Global Commun. Conf. (GLOBECOM)},
  2019, pp. 1--6.

\bibitem{hu2017potential}
S.~Hu, F.~Rusek, and O.~Edfors, ``The potential of using large antenna arrays
  on intelligent surfaces,'' in \emph{IEEE 85th Veh. Tech. Conf. (VTC Spring)},
  2017, pp. 1--6.

\bibitem{hu2018beyond}
------, ``{Beyond massive MIMO: The potential of data transmission with large
  intelligent surfaces},'' \emph{{IEEE} Trans. Signal Process.}, vol.~66,
  no.~10, pp. 2746--2758, 2018.

\bibitem{he2020adaptive}
J.~He, H.~Wymeersch, T.~Sanguanpuak, O.~Silv{\'e}n, and M.~Juntti, ``{Adaptive
  beamforming design for mmwave RIS-aided joint localization and
  communication},'' in \emph{IEEE Wireless Commun. Net. Conf. Work. (WCNCW)},
  2020, pp. 1--6.

\bibitem{jung2020asymptotic}
M.~Jung, W.~Saad, M.~Debbah, and C.~S. Hong, ``{Asymptotic optimality of
  reconfigurable intelligent surfaces: Passive beamforming and achievable
  rate},'' in \emph{IEEE Int. Conf. Commun. (ICC)}, 2020, pp. 1--6.

\bibitem{ozdogan2020using}
{\"O}.~{\"O}zdogan, E.~Bj{\"o}rnson, and E.~G. Larsson, ``{Using intelligent
  reflecting surfaces for rank improvement in MIMO communications},'' in
  \emph{IEEE Int. Conf. Acoustics, Speech and Signal Process. (ICASSP)}, 2020,
  pp. 9160--9164.

\bibitem{wu2019intelligent}
Q.~Wu and R.~Zhang, ``Intelligent reflecting surface enhanced wireless network
  via joint active and passive beamforming,'' \emph{{IEEE} Trans. Wireless
  Commun.}, vol.~18, no.~11, pp. 5394--5409, 2019.

\bibitem{wang2020intelligent}
P.~Wang, J.~Fang, X.~Yuan, Z.~Chen, and H.~Li, ``{Intelligent reflecting
  surface-assisted millimeter wave communications: Joint active and passive
  precoding design},'' \emph{IEEE Trans. Veh. Tech.}, 2020.

\bibitem{zhang2020reconfigurable}
H.~Zhang, B.~Di, L.~Song, and Z.~Han, ``{Reconfigurable intelligent surfaces
  assisted communications with limited phase shifts: How many phase shifts are
  enough?}'' \emph{IEEE Trans. Veh. Tech.}, vol.~69, no.~4, pp. 4498--4502,
  2020.

\bibitem{guo2019weighted}
H.~Guo, Y.-C. Liang, J.~Chen, and E.~G. Larsson, ``{Weighted Sum-Rate
  Maximization for Intelligent Reflecting Surface Enhanced Wireless
  Networks},'' in \emph{IEEE Global Commun. Conf. (GLOBECOM)}.\hskip 1em plus
  0.5em minus 0.4em\relax IEEE, 2019, pp. 1--6.

\bibitem{han2019large}
Y.~Han, W.~Tang, S.~Jin, C.-K. Wen, and X.~Ma, ``{Large intelligent
  surface-assisted wireless communication exploiting statistical CSI},''
  \emph{IEEE Trans. Veh. Tech.}, vol.~68, no.~8, pp. 8238--8242, 2019.

\bibitem{yang2020intelligentconf}
G.~Yang, X.~Xu, and Y.-C. Liang, ``{Intelligent reflecting surface assisted
  non-orthogonal multiple access},'' in \emph{IEEE Wireless Commun. Network.
  Conf. (WCNC)}, 2020, pp. 1--6.

\bibitem{zhang2020sum}
Y.~Zhang, C.~Zhong, Z.~Zhang, and W.~Lu, ``{Sum rate optimization for two way
  communications with intelligent reflecting surface},'' \emph{IEEE Commun.
  Lett.}, vol.~24, no.~5, pp. 1090--1094, 2020.

\bibitem{alegria2019achievable}
J.~V. Alegr{\'\i}a and F.~Rusek, ``{Achievable Rate with Correlated Hardware
  Impairments in Large Intelligent Surfaces},'' in \emph{IEEE Int. Workshop
  Computational Advances in Multi-Sensor Adaptive Process. (CAMSAP)}, 2019, pp.
  559--563.

\bibitem{gong2020towards}
S.~Gong, X.~Lu, D.~T. Hoang, D.~Niyato, L.~Shu, D.~I. Kim, and Y.-C. Liang,
  ``Towards smart wireless communications via intelligent reflecting surfaces:
  A contemporary survey,'' \emph{IEEE Commun. Surveys Tuts.}, 2020.

\bibitem{yang2020intelligent}
Y.~Yang, B.~Zheng, S.~Zhang, and R.~Zhang, ``{Intelligent reflecting surface
  meets OFDM: Protocol design and rate maximization},'' \emph{IEEE Trans.
  Commun.}, 2020.

\bibitem{yu2019miso}
X.~Yu, D.~Xu, and R.~Schober, ``{MISO wireless communication systems via
  intelligent reflecting surfaces},'' in \emph{IEEE Int. Conf. Commun. (ICCC)},
  2019, pp. 735--740.

\bibitem{yang2019irs}
Y.~Yang, S.~Zhang, and R.~Zhang, ``{IRS-enhanced OFDM: Power allocation and
  passive array optimization},'' in \emph{IEEE Global Commun. Conf.
  (GLOBECOM)}, 2019, pp. 1--6.

\bibitem{yuan2020intelligent}
J.~Yuan, Y.-C. Liang, J.~Joung, G.~Feng, and E.~G. Larsson, ``{Intelligent
  reflecting surface-assisted cognitive radio system},'' \emph{IEEE Trans.
  Commun.}, 2020.

\bibitem{di2020practical}
B.~Di, H.~Zhang, L.~Li, L.~Song, Y.~Li, and Z.~Han, ``{Practical Hybrid
  Beamforming With Finite-Resolution Phase Shifters for Reconfigurable
  Intelligent Surface Based Multi-User Communications},'' \emph{IEEE Trans.
  Veh. Tech.}, vol.~69, no.~4, pp. 4565--4570, 2020.

\bibitem{ying2020gmd}
K.~Ying, Z.~Gao, S.~Lyu, Y.~Wu, H.~Wang, and M.-S. Alouini, ``{GMD-based hybrid
  beamforming for large reconfigurable intelligent surface assisted
  millimeter-wave massive MIMO},'' \emph{IEEE Access}, vol.~8, pp.
  19\,530--19\,539, 2020.

\bibitem{landsberg2017design}
N.~Landsberg and E.~Socher, ``{Design and measurements of 100 GHz reflectarray
  and transmitarray active antenna cells},'' \emph{IEEE Trans. Antennas
  Propag.}, vol.~65, no.~12, pp. 6986--6997, 2017.

\bibitem{landsberg2017low}
------, ``{A low-power 28-nm CMOS FD-SOI reflection amplifier for an active
  F-band reflectarray},'' \emph{IEEE Trans. Microw. Theory Techn.}, vol.~65,
  no.~10, pp. 3910--3921, 2017.

\bibitem{bjornson_intelligent_2019}
E.~Bj\"{o}rnson, O.~\"{O}zdogan, and E.~G. Larsson, ``Intelligent reflecting
  surface vs. decode-and-forward: How large surfaces are needed to beat
  relaying?'' \emph{IEEE Wireless Commun. Lett.}, pp. 1--1, 2019.

\bibitem{di2020reconfigurable}
M.~Di~Renzo, K.~Ntontin, J.~Song, F.~H. Danufane, X.~Qian, F.~Lazarakis,
  J.~De~Rosny, D.-T. Phan-Huy, O.~Simeone, R.~Zhang \emph{et~al.},
  ``{Reconfigurable intelligent surfaces vs. relaying: Differences,
  similarities, and performance comparison},'' \emph{IEEE Open J. Commun.
  Society}, vol.~1, pp. 798--807, 2020.

\bibitem{pan2020multicell}
C.~Pan, H.~Ren, K.~Wang, W.~Xu, M.~Elkashlan, A.~Nallanathan, and L.~Hanzo,
  ``{Multicell MIMO communications relying on intelligent reflecting
  surfaces},'' \emph{IEEE Trans. Wireless Commun.}, 2020.

\bibitem{wu2020intelligent}
Q.~Wu, S.~Zhang, B.~Zheng, C.~You, and R.~Zhang, ``{Intelligent reflecting
  surface aided wireless communications: A tutorial},'' \emph{arXiv preprint
  arXiv:2007.02759}, 2020.

\bibitem{han2020cooperative}
Y.~Han, S.~Zhang, L.~Duan, and R.~Zhang, ``{Cooperative Double-IRS Aided
  Communication: Beamforming Design and Power Scaling},'' \emph{IEEE Wireless
  Commun. Lett.}, 2020.

\bibitem{zappone2013energy}
A.~Zappone, P.~Cao, and E.~A. Jorswieck, ``{Energy efficiency optimization in
  relay-assisted MIMO systems with perfect and statistical CSI},'' \emph{IEEE
  Trans. Signal Process.}, vol.~62, no.~2, pp. 443--457, 2013.

\bibitem{gong2019robust}
S.~Gong, S.~Wang, S.~Chen, C.~Xing, and L.~Hanzo, ``Robust energy efficiency
  optimization for amplify-and-forward mimo relaying systems,'' \emph{{IEEE}
  Trans. Wireless Commun.}, vol.~18, no.~9, pp. 4326--4343, 2019.

\bibitem{nguyen2019unequally}
N.~T. Nguyen and K.~Lee, ``{Unequally sub-connected architecture for hybrid
  beamforming in massive MIMO systems},'' \emph{IEEE Trans. Wireless Commun.},
  vol.~19, no.~2, pp. 1127--1140, 2019.

\end{thebibliography}

\end{document}